\documentclass[10pt,twocolumn]{article}
\usepackage{graphics}
\usepackage{mcite}
\usepackage{amsmath,amssymb}
\begin{document}

\title{Statistical mechanics of columnar DNA assemblies}
\author{A. Wynveen, D. J. Lee, and A. A. Kornyshev\\
Department of Chemistry, Imperial College London, SW7 2AZ London,
UK} \maketitle
 {\abstract{\bf Many physical systems can be mapped
onto solved or "solvable" models of magnetism. In this work, we
have mapped the statistical mechanics of columnar phases of
ideally helical rigid DNA - subject to the earlier found unusual,
frustrated pair potential [A. A. Kornyshev, S. Leikin, J. Chem.
Phys. 107, 3656 (1997)] - onto an exotic, unknown variant of the
XY model on a fixed or restructurable lattice. Here the role of
the 'spin' is played by the azimuthal orientation of the
molecules. We have solved this model using a Hartree-Fock
approximation, ground state calculations, and finite temperature
Monte Carlo simulations. We have found peculiar spin order
transitions, which may also be accompanied by positional
restructuring, from hexagonal to rhombohedric lattices. Some of
these have been experimentally observed in dense columnar
aggregates. Note that DNA columnar phases are of great interest in
biophysical research, not only because they are a useful in vitro
tool for the study of DNA condensation, but also since these
structures have been detected in living matter. Within the
approximations made, our study provides insight into the
statistical mechanics of these
systems.\\
  PACS:     {75.10.Hk}{ Classical spin models},
      {87.15.Nn}{ Properties of solutions; aggregation and
      crystallization of macromolecules},
      {64.70.-p}{ Specific phase transitions},
      {87.14.Gg}{ DNA,RNA}

}}
\section{Introduction}
\label{intro} DNA molecules in aqueous solution can be condensed
into a variety of phases. As the density of DNA is increased,
transitions occur from an isotropic liquid-like phase to a liquid
crystal phase and finally to a crystalline structure
\cite{livolant:96}. Within these phases, there exist
configurations with different symmetries and molecular
arrangements. For example, X-ray diffraction patterns of fibers of
different species of alkali metal salts of DNA reveal that the DNA
is crystallized into several different lattice types
\cite{langridge:60}. Likewise, liquid crystalline mesophases with
different symmetries, including cholesteric, line hexatic, and/or
hexagonal columnar phases, have been observed over a wide range of
DNA concentrations
\cite{livolant:96,robinson:61,strey:00,strzelecka:88}. These
mesophases are relevant for several reasons -- indeed they are
seen in many systems, such as bacteria, viruses, and
mitochondria\cite{reich:94}. Understanding the structure of DNA
aggregates may also aid in understanding the physics of DNA
packing into sperm and phage heads \cite{bloomfield:96} and gene
therapy vesicles \cite{gelbert:00,strey:98,podgornik:98}. Last but
not least, studies of these structures may shed light on the laws
of DNA-DNA interaction, important, e.g., in the problem of
recognition of homologous genes \cite{weiner:94}.
 As already mentioned, the specific phase of a DNA assembly depends
heavily on DNA concentration, but many other factors, such as
monovalent salt concentration and the effects of polycationic
condensing agents in the solution, will also greatly influence the
phase structure for a given DNA density
\cite{bloomfield:96,podgornik:98}. These many factors, along with
the complex chemical structure of DNA, complicate theoretical
studies of these phases and the transitions between them. In many
studies (see, for example, Ref.
\cite{oosawa:68,*stitger:77,*manning:78,*frank:87}) DNA molecules
are treated as uniformly charged cylinders since DNA is a
polyelectrolyte that dissociates in solution. This approximation
works well at large interaction distances where counterions may
screen the specific charge pattern of the DNA surface. But at the
smaller separations where liquid crystalline structures are
observed, a theory of the electrostatic interactions must take
into account the discrete helical structure of DNA. Recent
theoretical studies
\cite{kornyshev:97a,kornyshev:99a,kornyshev:98a,*kornyshev:98b}
have demonstrated that a number of phenomena can be rationalized
with the help of such a theory. Indeed, after dissociating in
solution, DNA preserves its double helical structure, with
negative charges residing on phosphate strands and specifically
adsorbing counter-cations settling in the grooves between the
phosphates. This results in a helical charge separation motif
along the DNA surface which dictates new interaction laws at close
range. Ref. \cite{kornyshev:97a} obtained a solution for the pair
potential between helical macromolecules in parallel alignment,
having in particular revealed that the interaction depends on the
relative azimuthal orientation of the molecules about their long
axes (negative charges on one molecule would like to be closer to
the positive charges on the other). Recent work has shown that
this azimuthal dependence yields a rich phase structure for
columnar aggregates and may explain distortions in the hexagonal
columnar phase \cite{harreis:02,*harreis:03,lorman:01}, as well as
a cholesteric to hexatic transition
\cite{kornyshev:00a,*kornyshev:02a}. The findings of Ref.
\cite{kornyshev:97a} allows a mapping of the pair interaction of
DNA onto an XY-spin model of magnetism, albeit with an unusual
spin coupling. In this report, we consider the effects of
temperature on the structure of columnar phases in order to
generate their full statistical mechanical description. Previous
work \cite{harreis:02,*harreis:03} obtained ground state
configurations for the columnar phases, but incorporating
temperature effects can change the properties of the transitions
between these phases. Furthermore, they may give rise to more
phases associated with the relative azimuthal orientation of the
DNA molecules.  We first develop a theory of columnar assemblies
fixed on a hexagonal lattice. We find additional
Berezinskii-Kosterlitz-Thouless-like and topologically-related
transitions in the 'spin'-structures when including temperature.
Also, for a more complete picture of the columnar assemblies, we
treat positional restructuring of the molecules at finite
temperatures. Incorporating spatial degrees of freedom into a
Monte Carlo simulation has yielded new insight into anomalous
spatial correlations in columnar phases, which may also have
relevance for transitions from columnar to cholesteric phases.
These studies reveal lattice types similar to those obtained for
the ground state \cite{harreis:02,*harreis:03} but great care must
be taken due to the limitations of the interaction potential for
certain situations. For example, at extremely large densities
where the DNA is closely packed, water, specifically its
temperature dependent dielectric constant, can no longer be
treated using its bulk properties as had been done to derive the
interaction potential.  Nevertheless, these results can be
considered valid over a wide range of physical parameters and act
as a first step to a completely atomistic approach which may
explore a broader range of the physically relevant parameter
space.

\section{Hexagonal columnar assemblies}
\label{sec:1} The calculation of the pair interaction between two
DNA in parallel juxtaposition \cite{kornyshev:97a} regarded DNA as
consisting of a double helical charge pattern of negative charges
along the phosphate spine of the DNA and positive charges adsorbed
into the grooves or on the phosphate strands (see Fig. 1). For
identical rigid helices, the ground state electrostatic
interaction energy between two DNA duplexes of length $L$ is given
by
\begin{equation}
\label{eq1}
E_{{\mathop{\rm int}} }  = L\left[ {a_0  - a_1 \cos
\left( {\phi _1  - \phi _2 } \right) + a_2 \cos \left( {2\left(
{\phi _1 - \phi _2 } \right)} \right)} \right]
\end{equation}
where the $a$ coefficients depend on a variety of factors such as
the charge distribution on the DNA and the dielectric properties
of the solution, and they decay exponentially with interaxial
spacing between the DNA \cite{kornyshev:97a,kornyshev:01a}; for
the most updated version of these expressions, see Appendix A of
Ref. \cite{lee:04}. $\phi_i$ characterizes the azimuthal
orientation of the middle of the minor groove of the $i$-th
molecule relative to the direction of interaxial separation (Fig.
1).

\begin{figure*}
\includegraphics[2cm,24cm][8cm,25.5cm]{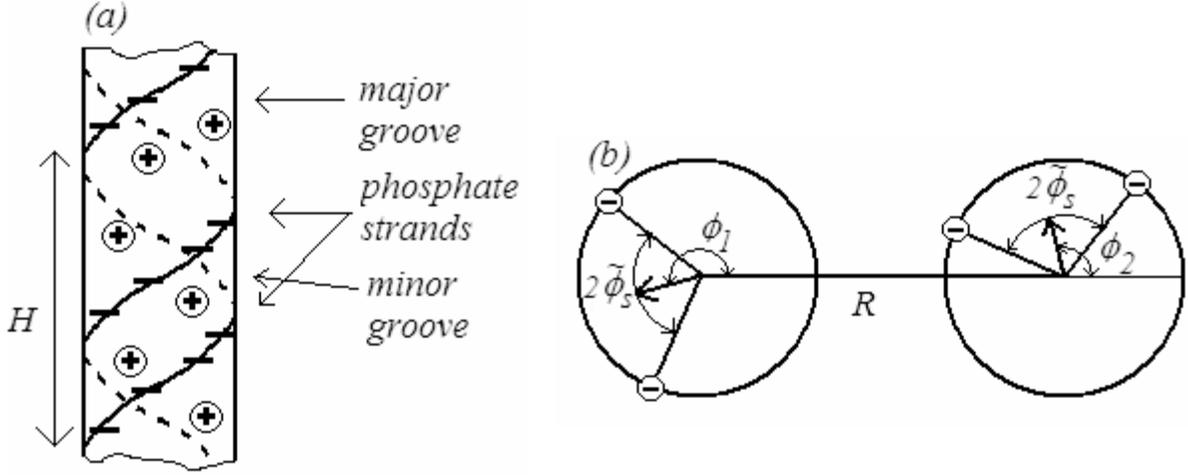}
\vspace*{5cm}       
 \caption{The charge distribution of a single
double helix (a) and the horizontal cross section of two identical
double helices in parallel juxtaposition (b), separated by
interaxial spacing $R$. The DNA double helix shown is considered
to consist of two spiralling negative phosphate strands with
specifically adsorbing cations in its minor and major grooves. The
pitch $H$ of the helix for B-DNA is approximately 34 $\AA$. In
(b), $\tilde \phi _s $ (which is about 0.4$\pi$ for B-DNA) is the
angular half-width of the minor groove between the phosphate
strands. 'Spins' characterize the azimuthal orientations of the
molecules; $\phi = \phi _1  - \phi _2$
 is the
angle of the relative azimuthal orientation of the molecules.}
\label{fig:1}       
\end{figure*}

At large interaxial separations, the $a_1$ term dominates the
$a_2$ term so the ground state energy is minimized when the two
duplexes have the same azimuthal orientations, {\it i.e.}, their
"spins" look in the same direction, $\phi  = \phi _1  - \phi _2  =
0$. But at separations below the point where $a_1  = 4a_2$, the
energy is minimized when $\phi  \ne 0$ and is degenerate:  $\phi =
\pm \phi _*$, where $\phi _*  = \arccos \left[ {a_1 /4a_2 }
\right]$. In a hexagonal lattice, this gives rise to frustration
between the neighboring spins which results in different spin
phases for the ground state, depending on the relative strength of
the $a$ coefficients.

Since these electrostatic coefficients decay rapidly for increased
spin separations, we need only consider nearest neighbor
interactions. The spin-dependent term in the Hamiltonian for a
hexagonal phase of rigid DNA fragments of length $L$ is given by
the 2D hexagonal lattice XY model where there is an additional
frustration term
\begin{align}
\label{ham}
  H &=  -{\raise0.5ex\hbox{$\scriptstyle 1$} \kern-0.1em/\kern-0.15em
\lower0.25ex\hbox{$\scriptstyle 2$}}La_1 \sum\limits_{ < ij> }
  {\cos \left( {\phi _i  - \phi _j } \right)}\nonumber \\
   &+ {\raise0.5ex\hbox{$\scriptstyle 1$} \kern-0.1em/\kern-0.15em
\lower0.25ex\hbox{$\scriptstyle 2$}}La_2 \sum\limits_{ < ij>}
   {\cos \left( {2\left( {\phi _i  - \phi _j } \right)}\right)}
\end{align}
with the summation over only nearest neighbors. The spin
configuration of the ground state for this Hamiltonian is
ferromagnetic, {\it i.e.}, all the spins are aligned, if $a_1  >
4a_2$. In the reverse case, the spins in the ground state are
aligned in a three-state Potts \cite{potts:52} type configuration
where the differences between the angles about any triangular
plaquette on the lattice possess values related
by\cite{cherstvy:04}
\begin{equation}
\label{eq3} \phi _2  - \phi _1  = \phi _1  - \phi _3  = \phi
_{potts}
\end{equation}
where
\begin{equation}
\label{eq4} \phi _{potts}  = \arccos \left[ {\frac{1}{4}\left( {1
+ \sqrt {1 + \frac{{2a_1 }}{{a_2 }}} } \right)} \right].
\end{equation}

Utilizing a self-consistent Hartree-Fock approximation (HFA)
\cite{samuel:82}, the free energy for a given spin configuration
can be obtained at finite temperatures. Derivations of these free
energies are left to the appendices. For the ferromagnetic state
where all the spins are aligned, the free energy is given by
\begin{align}
\label{Ffer} &F_{fer}  = \frac{{Nk_B T}}{2}\ln \left(
{\frac{J}{{k_B T}}}
\right) + Nk_B T\hat C_{Hex}  \nonumber\\
&+ 3NL\left[ {a_0  - a_1 \exp \left( { - \frac{{k_B T}}{{6J}}}
\right) + a_2 \exp \left( { - \frac{{2k_B T}}{{3J}}} \right)}
\right]
\end{align}
where $N$ is the total number of spins, $\hat C_{Hex}  =  - 1.265$
is a constant that depends on the geometry of the lattice, and $J$
is an effective coupling that is a solution to the transcendental
equation
\begin{equation} J = La_1 \exp \left( { - \frac{{k_B
T}}{{6J}}} \right) - 4La_2 \exp \left( { - \frac{{2k_B T}}{{3J}}}
\right).
\end{equation}
The free energy for the Potts state is more cumbersome and has the
form
\begin{align}
\label{Fpotts} F_{potts}  &= \frac{Nk_B T}{2}\ln \left(
\frac{J_1}{k_B T}
\right)\nonumber \\
&+ \frac{Nk_B T}{6}\tilde \Omega _{potts} (\alpha ) + Nk_B T\hat C_{Hex} \nonumber\\
&+ 2NL \left[ a_0  - a_1 \cos \left( \psi _{potts}
\right)\exp\left(
- \frac{\eta _1 (\alpha )k_B T}{J_1 } \right) \right. \nonumber\\
&\left. + a_2 \cos \left( 2\psi _{potts} \right)\exp \left( -
\frac{4\eta _1 (\alpha )k_B T}{J_1} \right) \right] \nonumber\\
&+ NL \left[ a_0  - a_1 \cos \left( 2\psi _{potts} \right)\exp
\left(  - \frac{\eta _2 (\alpha )k_B T}{J_1 } \right)\right.\nonumber \\
&\left.+ a_2 \cos \left( 4\psi _{potts}  \right)\exp \left(
 - \frac{4\eta _2 (\alpha ) k_B T}{J_1} \right) \right]
\end{align}
where $\tilde \Omega _{potts}$, $\eta_1$, and $\eta_2$ are given
in Appendix C and are functions of the ratio of the coupling
terms, $\alpha  = {{J_2 } \mathord{\left/
 {\vphantom {{J_2 } {J_1 }}} \right.
 \kern-\nulldelimiterspace} {J_1 }}$. This equation is closed by the additional transcendental
equations \addtocounter{equation}{+1}
\begin{align*}
J_1  &= La_1 \cos \left( {\psi _{potts} } \right)\exp \left( { -
\frac{{\eta _1 (\alpha )k_B T}}{{J_1 }}} \right)\\
&- 4La_2 \cos \left( {2\psi _{potts} } \right)\exp \left( { -
\frac{{4\eta _1 (\alpha )k_B T}}{{J_1 }}} \right)
 \tag{{\theequation}a}\label{j1potts}
\end{align*}
\begin{align*}
J_2  &= La_1 \cos \left( {2\psi _{potts} } \right)\exp \left( { -
\frac{{\eta _2 (\alpha )k_B T}}{{J_1 }}} \right)\\
 &- 4La_2 \cos
\left( {4\psi _{potts} } \right)\exp \left( { - \frac{{4\eta _2
(\alpha )k_B T}}{{J_1 }}} \right)\tag{{\theequation}b}
\end{align*}
\begin{align*}
&a_1 \sin \left( {\psi _{potts} } \right)\exp \left( { - \frac{{\eta _1 (\alpha )k_B T}}{{J_1 }}} \right)\\
&- 2a_2 \sin \left( {2\psi _{potts} } \right)\exp \left( { -
\frac{{4\eta _1 (\alpha )k_B T}}{{J_1 }}} \right) \\
 &+ a_1 \sin \left( {2\psi _{potts} } \right)\exp \left( { - \frac{{\eta _2 (\alpha )k_B T}}{{J_1 }}} \right)\\
 &- 2a_2 \sin \left( {4\psi _{potts} } \right)\exp \left( { - \frac{{4\eta _2 (\alpha )k_B T}}{{J_1 }}} \right) = 0
  \tag{{\theequation}c}.
 \end{align*}
To find the transition line between the ferromagnetic and Potts
states, the free energies (Eqs. (\ref{Ffer}) and (\ref{Fpotts}))
are equated after solving the corresponding transcendental
equations.

To obtain a full phase diagram, we also undertook Monte Carlo
studies of the interaction Hamiltonian, Eq. (\ref{ham}), on the
hexagonal lattice. The MC simulations were carried out in the
standard manner using the Metropolis algorithm. A lattice site was
chosen at random and the spin at this site was given a new random
orientation. The new interaction energy between this site and its
neighbors was calculated. If the new interaction energy was less
than that of the original state, the new spin orientation was
accepted. Otherwise, the new spin orientation was accepted with a
probability according to the Boltzmann factor $\exp \left( { -
\Delta E/k_B T} \right)$ where $\Delta E$ was the difference
between the new and old interaction energies. After equilibration
in the system was reached, successive MC steps were used to build
up the canonical distribution of the spin configurations to obtain
thermodynamic quantities such as specific heat, magnetic
susceptibility, etc.

\begin{figure*}
\includegraphics[2.5cm,24cm][8cm,25cm]{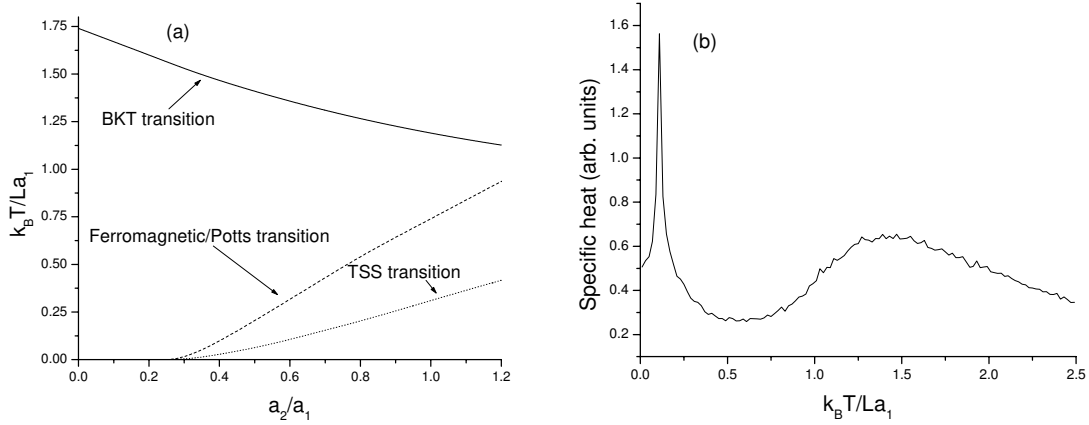}
\vspace{4.5cm} \caption{Location (a) of the various transitions in
a hexagonal lattice with nearest neighbor interactions of Eq.
(\ref{ham}). The TSS and BKT transitions are also observed as
peaks in the specific heat (b), here shown for a coupling ratio
$a_2/a_1=0.6$.}
\end{figure*}

In Fig. 2(a) we display the phase diagram of the system in terms
of the relative strengths of the $a$ coefficients and the
temperature. Also shown on this phase diagram is the location of
the transition between the Potts and ferromagnetic states obtained
from the analytical expressions of the free energies. The specific
heat for the ratio of the couplings ${a_2/a_1=0.6}$ is shown in
Fig. 2(b). The two peaks correspond to the transitions labeled as
the "TSS transition" and the "BKT transition" (defined below) in
the phase diagram.

The BKT-transition corresponds to the standard {\bf{B}}ere-zinskii
-{\bf{K}}osterlitz-{\bf{T}}houless transition observed in many 2D
spin systems \cite{kosterlitz:73}. At this transition, unbinding
of the vortices leads to an abrupt increase of the vortex density,
defined as $(1/N)\sum\limits_i {(v_i )^2 }$ where $v_i$ is the
vorticity at lattice site $i$, which in turn is defined as $\left(
{1/2\pi } \right)\sum\limits_\Delta  {\left( {\phi _i  - \phi _j }
\right)}$, the sum of the spin differences about a triangular
plaquette in the lattice. Figure 3 exhibits the increase in the
vortex density at this transition.

The peak in Fig. 2(b) found at lower temperatures stems from the
fact that there exist two topologically distinct ground states
\cite{top} of the Potts configuration (see Fig. 4(a)) akin to that
seen in purely antiferromagnetic spin systems\cite{lee:86}. In
Fig. 4(a) the number in the center of each triangular plaquette
corresponds to the positive helicity of that triangle, defined as
the sum of the clockwise positive change in the spin angles over
$2\pi$ as the triangle is traversed in the clockwise direction
(not to be confused with the vorticity where the angle difference
may have positive and negative values). This transition is
apparent in a plot of staggered, {\it i.e.}, taking into account
only downward-pointing triangles in the lattice, positive helicity
as the temperature changes (see Fig. 4(b)). Domain walls between
these two distinct topologies are excited destroying the staggered
helicity order at this transition, which we term the
{\bf{t}}opological {\bf{s}}pin {\bf{s}}tate (TSS) transition.

\begin{figure}
\includegraphics[2.5cm,24cm][5cm,25cm]{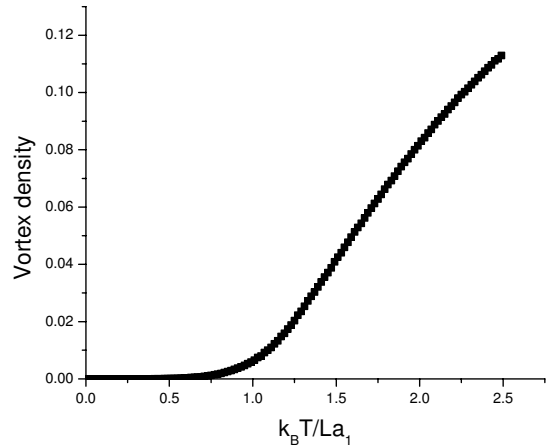}
\vspace{5.2cm} \caption{The vortex density as a function of
temperature for the same coupling ratio as in Fig. 2(b). At the
BKT transition, the vortex density begins to increase due to the
unbinding of vortices in the lattice.}
\end{figure}
\begin{figure*}
\includegraphics[2.5cm,24cm][5cm,25cm]{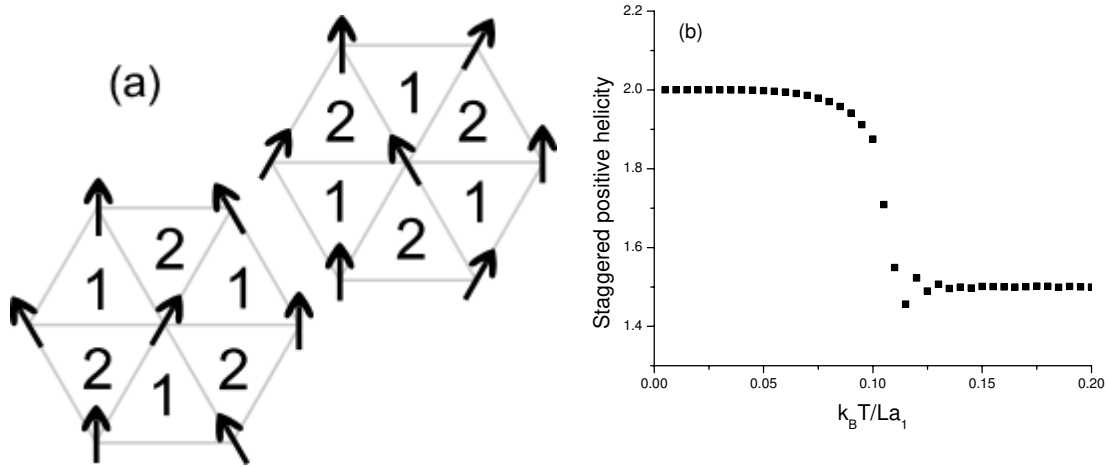}
\vspace{5cm} \caption{Two topologically distinct ground states for
the Potts configuration (a).  At the TSS transition, domain walls
between the two topologically distinct regions are excited so that
each state becomes equally probable. This can be seen in a plot of
the staggered positive helicity (defined in the text) as a
function of temperature (b), shown again for the coupling ratio of
Fig. 2(b).}
\end{figure*}

Comparisons can be made between the analytical forms, Eqs.
(5)-(8), and the MC simulations to confirm the validity of the two
methods. Specific heats of the ferromagnetic and Potts states can
be calculated from the free energy expressions. For the
ferromagnetic state in the region where $4a_2 /a_1  < 1$, there is
quite good agreement between the simulations and the analytic
forms (Fig. 5(a)).  As expected, deviations occur at higher
temperatures where the HFA breaks down. Agreement between
simulation and the analytical form for the Potts state, however,
is not as good (Fig. 5(b)). Here, the specific heat obtained from
the MC simulations diverges from the analytic form at relatively
low temperatures. This discrepancy, as well as the fact that the
Hartree approximation can not account for the phase transition,
can be taken as further indication of the TSS transition. The
Hartree approximation on its own neglects all topological
excitations \cite{top2}. Therefore, this difference starts being
significant at relatively low temperatures since domain walls may
be excited quite easily. Also shown in Fig. 5(c) is a high
temperature expansion of the specific heat for the
Kosterlitz-Thouless vortex state given by
\begin{eqnarray}
&\frac{{C_{KT} }}{{k_B }} = \frac{3}{2}\left( {\frac{{\left( {La_1
} \right)^2  + \left( {La_2 } \right)^2 }}{{\left( {k_B T}
\right)^2 }}} \right)\nonumber\\
 &+ 3\left( {\frac{{\left( {La_1 }
\right)^3 - \left( {La_2 } \right)^3 }}{{\left( {k_B T} \right)^3
}}} \right) - \frac{9}{{16}}\left( {\frac{{\left( {La_1 }
\right)^4  + \left( {La_2 } \right)^4 }}{{\left( {k_B T} \right)^4
}}} \right),
\end{eqnarray}
which conforms quite well to the simulations.

\begin{figure*}
\scalebox{1.2}{\includegraphics[3.2cm,24cm][8cm,25.5cm]{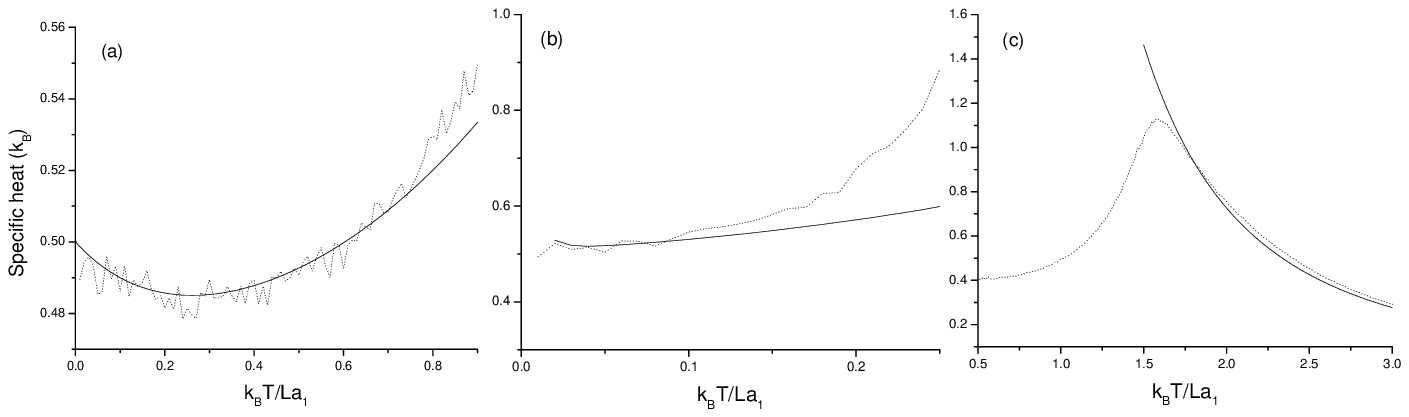}}
\vspace{4cm} \caption{Comparisons of the specific heat found from
analytical forms (solid lines) to that obtained from the MC
simulations (dotted lines). Results are displayed for the
ferromagnetic state (a) for a coupling ratio $a_2/a_1=0.1$ , the
Potts state (b) for a coupling ratio $a_2/a_1=1.0$, and finally
for the vortex regime (c) for a coupling ratio $a_2/a_1=0.2$.}
\end{figure*}

Now that we have constructed the phase diagram for the 2D
hexagonal lattice with spin interactions given by the Hamiltonian
of Eq. (\ref{ham}) for generic values of the  electrostatic $a$
coefficients, we can use their calculated values from Ref.
\cite{lee:04} to obtain results for 'real' DNA columnar phases.
Since these coefficients depend on the dielectric constant of the
solution, which is temperature dependent, the coefficients too
depend on temperature. Furthermore, interactions between DNA pairs
are screened by charges in the solution, and so these coefficients
are functions of the effective inverse Debye screening length. In
a polyelectrolyte assembly with Donnan equilibrium
\cite{cherstvy:02}, this is given by \cite{harreis:02,*harreis:03}
\begin{equation}
\kappa  = \sqrt {4\pi \frac{{\left( {Z\rho /L + 2n_s } \right)e^2
}}{{\varepsilon k_B T}}}
\end{equation}
and so there is an additional temperature dependence in the
coefficients.  Here, $e$ is the electron charge, $n_s$ is the salt
concentration, $\rho$ is the 2D DNA density in the columnar
assembly, and $Z\left| e \right|$ is the uncompensated DNA charge
(the fraction of the negative phosphate backbone that is not
compensated by readsorbed cations). After finding the coefficients
over a range of relevant temperatures and densities for a given
charge distribution on the DNA and salt concentration, we simply
map these coefficients onto the phase diagram of Fig. 2(a) at the
corresponding temperature to find where the transitions occur.

For ambient conditions at which columnar phases are observed {\it
in vivo} or {\it in vitro}, the BKT transition would unlikely be
observed. The relative strength of thermal energy to the
interaction energy, which is on the order of unity at the BKT
transition, would engender that extremely high, physically
unviable, temperatures would be necessary to reach this
transition. Rather, instead of going to high temperatures, the
electrostatic interactions could be weakened, namely by increasing
the salt concentration or increasing the interaxial distances
between the DNA molecules, so that thermal energies would be
comparable to the interaction energy. But at such low densities,
the DNA would no longer be in a columnar aggregate.  Likewise, at
large salt concentrations (approximately ten times that of
physiological levels) where this transition could be seen, the
interaction between DNA would be so weak that we could not assume
the DNA are pinned to the hexagonal lattice sites. In the next
section, we will include the effects of thermally induced
positional restructuring in the lattice showing that such effects
certainly cannot be neglected.

On the other hand, the TSS transition may occur at temperatures,
DNA densities, and salt concentrations where columnar structures
are observed in experiments. At this transition, the electrostatic
interaction energy dominates the thermal energy of the DNA so that
the DNA is essentially immobile within the assembly. In Fig. 6 we
show the location of this transition for DNA with different charge
compensation, {\it i.e.}, the fraction of the negative charge on
the DNA phosphate backbone that is compensated by readsorbed
cations, and also different salt concentration in the solution
\cite{kornyshev:99a}. For all of these cases, we assumed that 30\%
of the cations readsorbed into the minor groove and 70\% into the
major groove of the DNA. To obtain this phase diagram, we took the
length of the DNA molecules to be $L_p=50 \mbox{nm}$, which is the
persistence length of DNA.

We must be careful however since this transition occurs at
relatively small interaxial spacings where the continuum
electrostatic theory, which underlies the Hamiltonian of Eq.
(\ref{ham}), may break down. Although these spacings are quite
small, they are close to the Debye screening length and so the
linearized Poisson-Boltzmann equation used to derive the
interaction \cite{kornyshev:97a} may still be valid at first
approximation. However, dielectric response of water in the
confined intracolumnar space may be different than in the bulk:
most likely the effective dielectric constant of water will be
reduced there. Its variation from the bulk value will vary with
the density of the aggregate. This would slightly shift the
transition to larger interaxial spacings, because of an increase
in the electrostatic interactions between the DNA but would not
eliminate this transition altogether.

\begin{figure}
\includegraphics[2.5cm,24cm][5cm,25.5cm]{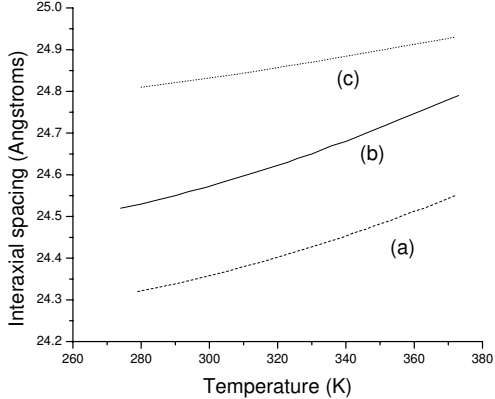}
\vspace{4.7cm} \caption{Location of the TSS transition for
different charge distributions on the DNA and different salt
concentrations in the solution. The topologically disordered state
lies above the curves, corresponding to lower densities. The
dashed line (a) corresponds to a charge compensation of 0.9 with
0.1 M salt concentration.  The solid line (b) and dotted line (c)
correspond to a charge compensation of 0.7 with 0.1 M salt
solution for (b) and 1.0 M for (c). The effects are non-trivial,
as they are driven mainly by the variation of $\kappa$ (Eq.(10))
which changes the balance between the electrostatic coefficients
$a_1$ and $a_2$.}
\end{figure}


\section{Positional restructuring of the assembly}
 \label{sec:2}
So far, we have assumed that the DNA molecules are pinned to sites
on a hexagonal lattice. However, thermal motion, especially when
the interactions grow weaker, may distort the lattice and, hence,
affect the statistical mechanical properties of the system.
Furthermore, depending on the form of the interaction, namely the
interplay among the $a$ coefficients, the hexagonal lattice may
not be the optimal ground state of the system. Previous
experimental and theoretical works have demonstrated that the 2D
lattice of a columnar assembly may be distorted hexatic
\cite{strey:00,lorman:01} or, almost equivalently, rhombic
\cite{harreis:02,*harreis:03}. To incorporate these effects, we
must modify our model for the columnar phases of the previous
section.

To build up the statistical mechanical theory for columnar
assemblies including positional restructuring, the ground state
configuration of the system is first obtained by performing a
lattice sum of the interactions among the DNA molecules in the
assembly \cite{harreis:02,*harreis:03}. Upon carrying out the
lattice summation, we find that the ground state configuration for
certain values of the electrostatic coefficients is no longer
hexagonal but rather rhombic as seen in Ref.
\cite{harreis:02,*harreis:03}. For this structure, the spins
possess a {\bf{q}}uasi-{\bf{a}}nti{\bf{f}}erro-magnetic-Ising
(QAF) ordering: spins at opposite diagonals of the rhombic cell
are the same but there is a finite difference between the spins at
adjacent corners (see Fig. 7). This is quite understandable
qualitatively: parallel spins will repel each other due to the
frustration-inducing $a_2$ interaction term thus distorting the
hexagonal lattice. The degree of this distortion again rests on
the interplay of the $a$ terms in the interaction.

\begin{figure}
\includegraphics[2.5cm,24cm][5cm,26cm]{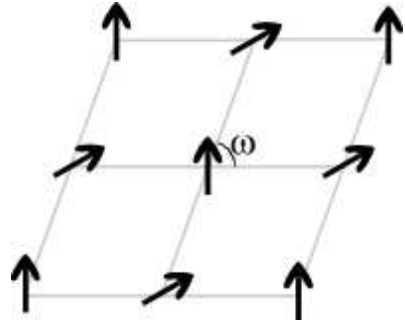}
\vspace{3.3cm} \caption{The rhombic QAF state showing the
distortion from the hexagonal lattice ($ \omega  \ge 60^ \circ$).
Parallel spins lie at opposite corners of the rhombi whereas spins
with parameter-dependent differences lie at adjacent corners.}
\end{figure}
As before, we have derived an analytical expression for the free
energy for this configuration taking into account only the six
nearest neighbors. Here, however, due to the distortion, the
electrostatic coefficients each take on two values according to
the direction taken along the lattice: four of the nearest
neighbors are at adjacent corners and the remaining two at the
opposite corners along the short diagonal. Note, that this
neglects neighbors across the long diagonal, but excluding these
can be justified as long as the distortion angle is not much
larger than $60^ \circ$, which is generally the case.  The free
energy for this state is given by {\scriptsize
\begin{align}
&F_{af}  = \frac{{Nk_B T}}{2}\ln \left( {\frac{{J_1 }}{{k_B T}}}
\right)
 + \frac{{Nk_B T}}{6}\tilde \Omega _{af} (\alpha
+ \frac{1}{2}) + Nk_B T\hat C_{Hex}\nonumber\\
 &{\rm  + }2NL\left[ {a_0 (R_1 )
- a_1 (R_1 )\cos \left( {\psi _{af} } \right)\exp \left( { -
\frac{{k_B T}}{{\pi J_1 }}\arcsin \frac{1}{{\sqrt {2\left( {\alpha
+ 1} \right)} }}} \right)} \right.\nonumber\\
&\left. {{\rm + }a_2 (R_1 )\cos \left( {2\psi _{af} } \right)\exp
\left( { - \frac{{4k_B T}}{{\pi J_1 }}\arcsin \frac{1}{{\sqrt
{2\left({\alpha  + 1} \right)} }}} \right)} \right]\nonumber\\
  &{\rm  + }NL\left[ {a_0 (R_2 ) - a_1 (R_2 )\exp \left( { - \frac{{k_B T}}
  {{2J_2 }}\left( {1 - \frac{4}{\pi }\arcsin \frac{1}{{\sqrt
  {2\left( {\alpha  + 1} \right)} }}} \right)} \right)}
  \right.\nonumber\\
  &\left. {{\rm + }a_2 (R_2 )\exp \left( { - \frac{{2k_B T}}{{J_2 }}
  \left( {1 - \frac{4}{\pi }\arcsin \frac{1}{{\sqrt {2\left(
  {\alpha  + 1} \right)} }}} \right)} \right)} \right]
\end{align}}\normalsize
where $\alpha  = J_2 /J_1$, the form of $\tilde \Omega _{af}$ is
given in Appendix B, and the $a$ coefficients are calculated at
interaxial separations of $R_1$, the distance between differing
spins, and $R_2  = R_1 \sqrt {2 - 2\cos \omega }$, the distance
between parallel spins across the short diagonal. The equation is
closed by the additional equations \addtocounter{equation}{+1}
{\footnotesize
\begin{align*}
&J_1  = La_1 (R_1 )\cos \left( {\psi _{af} } \right)\exp \left( { - \frac{{k_B T}}{{\pi J_1 }}\arcsin \frac{1}{{\sqrt {2\left( {\alpha  + 1} \right)} }}} \right) \\
&{\rm     } - 4La_2 (R_1 )\cos \left( {2\psi _{af} } \right)\exp
\left( { - \frac{{4k_B T}}{{\pi J_1 }}\arcsin \frac{1}{{\sqrt
{2\left( {\alpha  + 1} \right)} }}} \right) \tag{{\theequation}a}
\end{align*}
\begin{align*}
&J_2  = La_1 (R_2 )\exp \left( { - \frac{{k_B T}}{{2J_2 }}\left( {1 - \frac{4}{\pi }\arcsin \frac{1}{{\sqrt {2\left( {\alpha  + 1} \right)} }}} \right)} \right) \\
 &{\rm     } - 4La_2 (R_2 )\exp \left( { - \frac{{2k_B T}}{{J_2 }}\left( {1 - \frac{4}{\pi }\arcsin \frac{1}{{\sqrt {2\left( {\alpha  + 1} \right)} }}} \right)} \right)
\tag{{\theequation}b}
\end{align*}
\begin{equation*}
\cos \left( {\psi _{af} } \right) = \exp \left( { - \frac{{3k_B
T}}{{\pi J_1 }}\arcsin \frac{1}{{\sqrt {2\left( {\alpha  + 1}
\right)} }}} \right)\frac{{a_1 (R_1 )}}{{4a_2 (R_2
)}}.\tag{{\theequation}c}
\end{equation*}}\normalsize

\begin{figure}
\scalebox{1.2}{\includegraphics[3.8cm,23cm][5cm,25.cm]{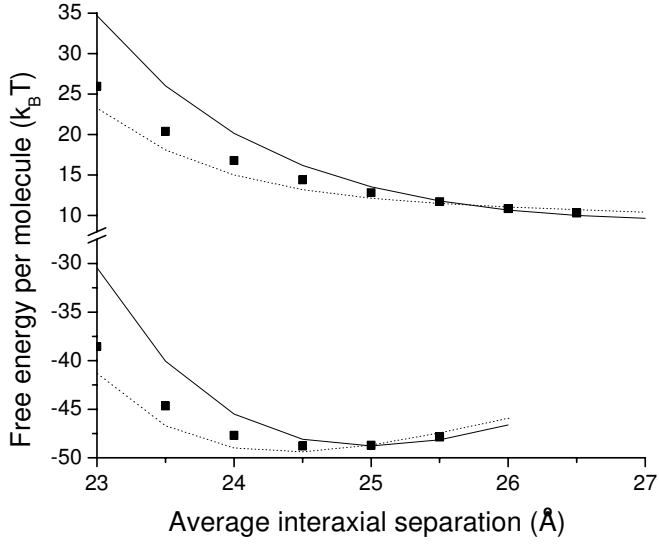}}
\vspace{5cm} \caption{The free energy for the QAF rhombic state
(dotted line), Potts configuration (squares) and ferromagnetic
hexagonal states (solid line) as a function of the average
interaxial separation between nearest neighbors.  The data are
given for one persistence length long DNA and with charge
distributions on the surface of 30\% readsorbed cation charge in
the minor groove and 70\% in the major groove for a charge
compensation of 0.9 (below the break in the y-axis) and of 0.7
(above the break). For a charge compensation of 0.9, the QAF
rhombic state is the minimum energy configuration below a
separation of about 25.0$\AA$ [the Potts state is the lowest
energy configuration at smaller spacings (not shown)], and the
ferromagnetic hexagonal state is the favorable configuration above
this point. Likewise, for a DNA charge compensation of 0.7, this
transition occurs at a separation of about 25.7$\AA$.}
\end{figure}

Solving Eqs. (11) and (12), we can find the amount of distortion
that minimizes the free energy. Figure 8 shows comparisons between
the analytic forms for the free energy at a temperature of 300 K
for the various lattice types for a couple of different charge
distributions on the DNA surface. Incorporating temperature via
the HFA calculation leads to an abrupt first order transition
between the QAF rhombic state and the ferromagnetic hexagonal
state, found from comparing the results of Eq. (5) and Eq. (11).
This may well be an artifact of the HFA calculation \cite{lee:04},
but nevertheless, the transition becomes much sharper as compared
to the smoother crossover found from the ground state lattice sum
calculation. This transition also occurs at a smaller interaxial
spacing than where the ground state lattice sum calculation yields
the crossover between the two states. Including temperature also
alters the amount of the distortion in the rhombic state, as shown
in a plot of $\omega$ as a function of average interaxial spacing
in Fig. 9, as compared to that found from the lattice sum
calculations. At very large densities, the system returns once
again to a hexagonal state with the Potts-type spin configuration.
Also, as seen in Fig. 8, the energy difference between the
hexagonal Potts configuration and the QAF rhombic state is on the
order of $k_BT$, and so there may be a mixture of these phases at
certain densities. Again, these results are subject to the
limitations of the derived pair interaction at small interaxial
spacings mentioned previously.

\begin{figure}
\includegraphics[3cm,24cm][5cm,25.cm]{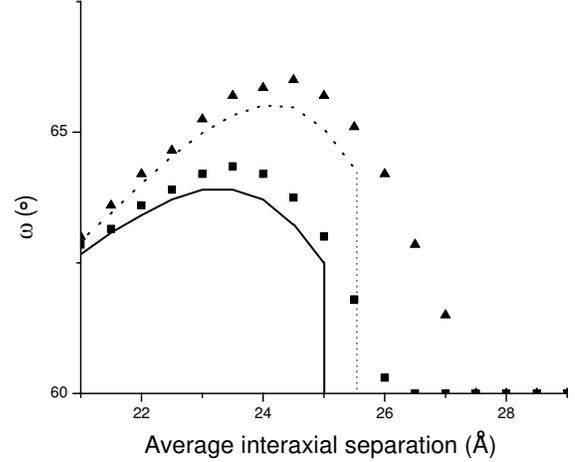}
\vspace{5.2cm} \caption{Comparisons of the level of distortion
(values of the rhombic angle $\omega$) obtained from the analytic
forms (lines) and the ground state lattice summation (symbols) as
a function of average interaxial separation. Results are shown for
a DNA charge compensation of 0.9 (solid line and squares) and of
0.7 (dotted line and triangles).  The transition from the QAF
rhombic state to the ferromagnetic state is of first order in the
HFA calculation and occurs at smaller interaxial spacings as
compared to the smooth crossover obtained from ground state
lattice sum calculations.}
\end{figure}

As before, we perform Monte Carlo simulations on the 2D columnar
system. As opposed to the fixed lattice calculations of the
previous section, here, however, a MC step either corresponds to a
change in the spin or a change in position to probe both spin and
spatial degrees of freedom. Again, the standard Metropolis
algorithm is employed to build up the canonical distribution of
the assembly at a specified temperature. The new energy is found
by calculating the interactions of the molecule only with others
that lie within a specified neighborhood of the original molecule.
This is done to increase the efficiency of the simulation and is
justified since the interactions decay quickly with increasing
intermolecular distance. From the simulations, we obtain relevant
thermodynamic quantities, upon thermal equilibration of the
system, as well as information concerning spin and positional
correlations.

The QAF rhombic to ferromagnetic hexagonal transition shown in
Figs. 8 and 9 also appears in these simulations at the same
average nearest neighbor interaxial spacings found from our
analytical explorations. In Fig. 10, the distribution of the spin
difference between nearest neighbors is shown for interaxial
spacings just below and above this transition.  As shown in this
figure, the nearest neighbor spin configuration develops from the
two-state QAF phase to a broad single maximum about zero, the
ferromagnetic phase. This is likewise observed in the spatial
distribution between neighbors. Below the transition, the
distribution shows nearest neighbor, next nearest neighbor, etc.
correlations that match a rhombic structure, while above the
transition, the distribution matches that of a hexagonal lattice.
Note that just beyond this transition at lower densities (at least
for the charge distribution on the DNA used in the simulation that
generated the results of Fig. 10), the DNA precipitates out of
solution leaving a coexistence region of DNA aggregate and
DNA-free solution \cite{harreis:02,*harreis:03}.

\begin{figure}
\includegraphics[3cm,24cm][5cm,25cm]{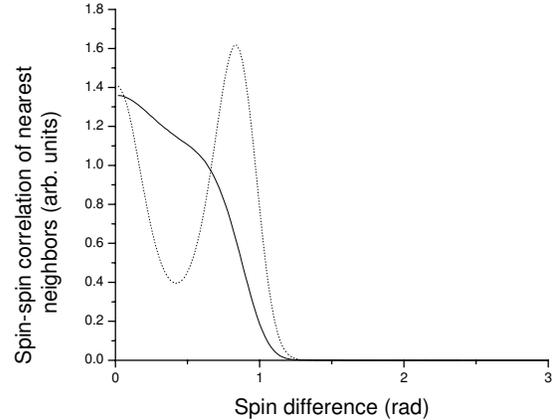}
\vspace{5cm} \caption{The thermally averaged probability
distribution of the spin difference between nearest neighbors near
the transition between the QAF rhombic state and the ferromagnetic
hexagonal state.  The distributions are shown for a DNA charge
compensation of 0.9 with the dotted line corresponding to an
interaxial spacing of 24.5$\AA$ and the solid line to the one  of
25.4$\AA$ at a temperature of 300K.}
\end{figure}

In order to avoid the situation where DNA condenses out of
solution at lower densities solely due to the choice of the charge
distribution on the DNA, a charge distribution can be used where
the interaction energy leads to a purely repulsive force between
molecules.  One such charge distribution is that where there the
readsorbed cations are shared evenly between the major and minor
grooves of the double helix. For this charge distribution, the
interaction energy found by lattice sum calculations yields a
flattening of the potential at intermediate densities that arises
due to the spin frustration in the system.  Due to this
flattening, we find that as the density is increased, the spatial
correlation function (the probability distribution of the location
of neighboring molecules, $4\pi R^2g(R)$ ) becomes more liquid
like. Results of the MC simulation are shown in Fig. 11
demonstrating this effect. As the density is increased further, of
course, the system once again becomes more crystalline. We
essentially find here an effect which could be conventionally
called spin-frustration induced melting of the positional
structure of the columnar aggregate.

\begin{figure}
\scalebox{1.1}{\includegraphics[3.5cm,24cm][5cm,25cm]{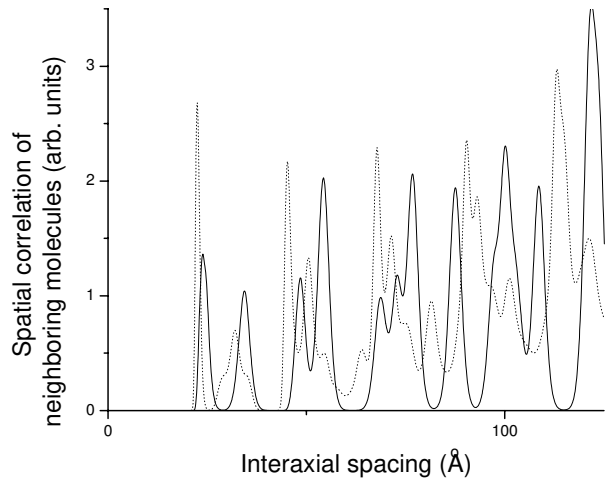}}
\vspace{5.5cm} \caption{The probability distribution of neighbors
for a DNA surface charge distribution with 90\% charge
compensation where 50\% of the cations are readsorbed in the minor
groove and 50\% in the major groove.  The dotted line corresponds
to an average interaxial nearest neighbor spacing 24.2$\AA$ and
the solid line corresponds to an average spacing of  26.2$\AA$. As
evident in the plot, the lower density distribution is more
liquid-like where neighboring molecules have a finite probability
to be at a broad range of interaxial spacings whereas the lower
density spatial distribution possesses a more crystalline
ordering.}
\end{figure}

\begin{figure*}
\includegraphics[2.5cm,24cm][8cm,26cm]{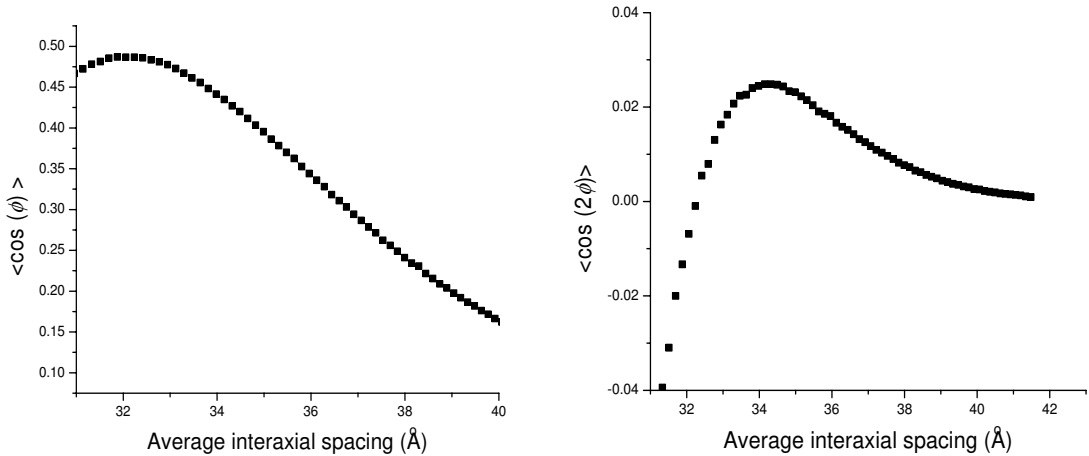}
\vspace{5cm} \caption{Biaxial correlation functions calculated
from the probability distribution of nearest neighbor spin
differences.}
\end{figure*}

In Ref. \cite{kornyshev:00a,*kornyshev:02a} analysis of the chiral
electrostatic interaction for cholesteric liquid crystals
demonstrated that biaxial correlations, {\it i.e.}, the
ensemble-averaged azimuthal spin difference between nearest
neighbors, determines the strength of the chiral interaction in
the cholesteric phase. As these correlations grow larger, the
chiral interaction, giving rise to the cholesteric phase,
increases.  These correlations, $\left\langle {\cos (\phi )}
\right\rangle$ namely and $\left\langle {\cos (2\phi )}
\right\rangle$, can easily be calculated from the thermally
averaged distribution of the spin difference between nearest
neighbors in our simulation. As shown in Fig. 12, $\left\langle
{\cos (\phi )} \right\rangle$ obtains its maximum value at an
average interaxial separation of about 32$\AA$ at the same point
that $\left\langle {\cos (2\phi )} \right\rangle$ becomes
positive. Interestingly, this spacing corresponds to that at which
a transition between the columnar phase and the cholesteric phase
is observed in experiments \cite{strzelecka:87,durand:92}. The
transition between the QAF rhombic and ferromagnetic hexagonal
phases, as shown in Fig. 9 for different charge distributions,
correspondingly occurs at this same spacing for this DNA charge
distribution. As the interaxial spacing increases, these
correlations decay to zero as a result of the overall weakening of
the interactions between molecules at smaller densities, which
leads to an isotropic liquid phase.

%
%

\section{Discussion}
\label{discussion} We have presented results of a statistical
mechanical analysis of DNA in columnar assemblies that interact
via an azimuthal angle ('spin') dependent pair potential
\cite{kornyshev:97a}. Initially, we assumed that DNA is packed in
a fixed two dimensional hexagonal structure so that we had to
consider only 'spin-spin' interactions between neighboring
molecules. Hence, tools similar to those used for studying
magnetic systems could be employed. Again, this spin interaction
arises from the helical charge distribution on the double helix.
Due to a quasi-antiferromagnetic coupling term in the interaction,
the system may be frustrated, which gives rise to a rich phase
structure.

Besides a Berezinskii-Kosterlitz-Thouless type transition and an
intermediate transition to a ferromagnetic-like state, we have
found a transition associated with domain formation of two
distinct topologies of the spin system. Comparing biologically
relevant values of the spin coupling terms to the generic phase
diagram of the hexagonal system, we find that this latter
transition may be experimentally probed in dense assemblies.
Calorimetric measurements would seem to be the simplest option to
study this transition, but other experimental factors may inhibit
a direct measurement of the transition in this manner. X-ray
diffraction \cite{strey:00} or NMR \cite{strzelecka:87}
techniques, however, may be able to directly scrutinize the spin
structure within the assembly to obtain the spin correlations in
the system thus providing evidence if such a transition exists.
Investigations are currently being pursued by our experimental
collaborators at Imperial.

Extending the theory, we considered non-hexagonal lattice types
and also incorporated thermally-induced positional fluctuations of
the molecules. Previous studies of DNA assemblies
\cite{harreis:02,*harreis:03} demonstrated that a 2D rhombic
(distorted hexagonal) lattice would be the ground state
configuration of the system under certain conditions. We found
from both analytical and numerical investigations a first order
transition from this quasi-antiferromagnetic rhombic state to a
hexagonal ferromagnetic state as the density of DNA in the system
decreases. Likewise, this transition also appears in the
probability distribution of the nearest neighbor phase angle
difference of the MC simulations. The ensemble average of nearest
neighbor spin differences develops from a two-state spin system to
one with a single maximum.

Furthermore, these MC simulations revealed other interesting
phenomena. For certain distributions of adsorbed cations on the
DNA surface, we found that increasing the DNA density lead to a
counterintuitive reduction in the crystalline ordering of the
system so that the system became more liquid like. This has been
observed in classical experiments\cite{strey:00} of densely packed
DNA at roughly the same interaxial spacings as those found in our
simulations. Again, this results from the frustration in the spin
interaction between the molecules. Also, in this same system, we
found that the behavior of biaxial correlations between
neighboring molecules, which has been proposed as a mechanism
underlying the formation of cholesteric phases
\cite{kornyshev:00a,*kornyshev:02a} at densities lower than those
of the columnar phases, is influenced by the transition between
the quasi-antiferromagnetic rhombic and ferromagnetic hexagonal
phases. This occurs at interaxial spacings where the transition
between columnar and cholesteric phases are observed in
experiments. As the density of the DNA-assembly is further
decreased, the intermolecular interactions would grow ever weaker,
until, finally, thermal motion would destroy any lattice ordering
in the system. At this point, the lattice would then completely
melt into a liquid-like isotropic phase \cite{kassapidou:98}.

Throughout this paper we have used a 2D model to describe
interactions between DNA.  In three dimensions, as well as
azimuthal angular fluctuations, there exist fluctuations of the
relative positions of the molecules along their long axes.  If the
charge distribution of the DNA surface is taken as continuous,
then these two types of fluctuations are indistinguishable in the
roles they play in the interaction energy, {\it i.e.}, an
azimuthal rotation is equivalent to a translation along the long
axis. For this case, we may simply replace the azimuthal
coordinate $\phi$ by the coordinate $\tilde \phi  = \phi  - {{2\pi
z} \mathord{\left/
 {\vphantom {{2\pi z} H}} \right.
 \kern-\nulldelimiterspace} H}$
in all our expressions, which would not alter our present results.
However, if we consider that the charges on the DNA surface are
discrete, this equivalence will be lost; there will be additional
interactions which destroy this symmetry. Nevertheless, if the
assemblies are not extremely dense, these additional interactions
due to discreteness can be neglected and the charge pattern on the
DNA can be considered to be continuous. Still, discreteness will
restrict the corkscrew motion ($\tilde \phi  = 0$) of DNA, which
costs no energy for a continuous charge distribution, within the
assemblies.

We must caution once again that the form of the DNA-DNA
interaction \cite{kornyshev:97a} we have employed is subject to
certain limitations. As the concentration of DNA in the solution
is increased, so that the surface-to-surface separation becomes
prohibitively small, effects of nonlocal polarizability of the
water in the narrow interstitial regions between the DNA could
alter the results \cite{solvation}. Furthermore, dielectric
saturation and sterical constraints threaten to freeze the
dielectric response of strongly confined water. Likewise, at such
large densities, steric forces between DNA mediated by confined
water may need to be also included in the interaction. Also, the
pair potential was derived using a linearized Poisson-Boltzmann
equation which is valid for weak electric fields in solution, and
thus the pair potential may not be valid when the electrostatic
interactions are quite substantial (large densities).  Next, as
the DNA concentration is altered or the DNA move about,
rearrangement or additional adsorption/desorption of cations on
the DNA surface may occur in response to a change in
intermolecular distances \cite{cherstvy:02}, thus affecting the
pair interaction. Last but not least, in this study DNA duplexes
have been considered as ideally helical: sequence dependent
distortions from an ideal double helical step structure of DNA
\cite{kornyshev:01a} has been neglected here as well as the
torsional elasticity of the duplexes, that in columnar aggregates
can correct the nonideality
\cite{lee:04,cherstvy:04,kornyshev:04a}. Nevertheless, this is a
first step to rationalize the columnar systems at finite
temperatures before treating them with a computationally expensive
fully atomistic approach in which these results may be tested.

\section{Acknowledgements}
\label{acknowledgements} This work has been strongly stimulated by
the Monte Carlo simulations of DNA columnar assemblies of G.
Sutmann and many detailed discussions of this subject with him and
with also S. Leikin. Financial support from EPSRC grant N:
GR/S31068/01 and a Royal Society Wolfson Merit Research award to
A.A.K. are gratefully acknowledged.

\renewcommand{\theequation}{A\arabic{equation}}
\setcounter{equation}{0}
\section*{Appendix A: The Hartree approximation for the
ferromagnetic state} \label{appa} The partition function for our
modified XY model is
\begin{equation}
Z = \prod\limits_{jl} {\int {d\phi _{j,l} } }\; {\mathop{\rm
exp}\nolimits} \left( { - \frac{{E[\phi _{j,l} ]}}{{k_B T}}}
\right)
\end{equation}
where we may rewrite Eq. (2) of the text as
\begin{equation}
\begin{array}{l}
 E[\phi _{j,l} ] = L\sum\limits_{p = 1}^2 {} \sum\limits_{j,l} {a_p ( - 1)^p [\cos (p(\phi _{j,l}  - \phi _{j,l - 1} ))}  \\
 {\rm            } + \cos (p(\phi _{j,l}  - \phi _{j - 1,l} )) + \cos (p(\phi _{j,l}  - \phi _{j - 1,l + 1} ))]
 \end{array}.
 \end{equation}
Here, we have introduced two lattice vectors $\vec u_i  = jr_0
\hat u$ and $\vec v_j  = lr_0 \hat v$, where $\hat u$ and $\hat v$
are unit vectors, to describe the relative positions of the sites
(DNA molecules) in the hexagonal lattice. The relative positions
of these two vectors, as well as site labelling, are shown in Fig.
13.
\begin{figure}
\includegraphics[2.5cm,24cm][5cm,22.cm]{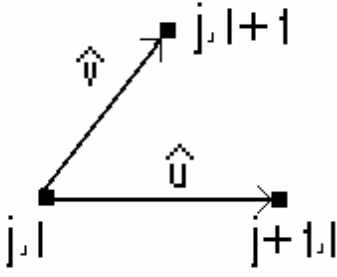}
\vspace{4cm} \caption{Lattice labelling and relative positions of
$\hat u$ and $\hat v$.}
\end{figure}
We then start by expanding out $E[\phi_{j,l}]$ in powers of $\phi$
and dividing the energy into a Gaussian part as well as an
anharmonic part, which contains terms of higher order than two in
the expansion.
\begin{equation*}
E[\phi _{j,l} ] = E_0 [\phi _{j,l} ] + E_{anh} [\phi _{j,l} ]
\end{equation*}
\begin{align}
E_0 [\phi _{j,l} ] &= 3NL(a_2  - a_1 ) + \frac{{m_0
}}{2}\sum\limits_{j,l} \left[ {(\phi _{j,l}  - \phi _{j - 1,l} )^2
}\right.\nonumber\\
&\left.+ (\phi _{j,l}  - \phi _{j,l - 1} )^2  + (\phi _{j,l} -
\phi _{j - 1,l + 1} )^2\right]
\end{align}
{\scriptsize
\begin{align}
&E_{anh} [\phi _{j,l} ] = \sum\limits_{j,l} {\sum\limits_{n =
2}^\infty } \frac{1}{(2n)!}\left[\left( {La_1 ( - 1)^{n - 1} +
La_2 ( - 4)^n } \right)\right.\nonumber\\
&\left. \times \left( (\phi _{j,l}  - \phi _{j - 1,l} )^{2n}
 + (\phi _{j,l}  - \phi _{j - 1,l} )^{2n}
+ (\phi _{j,l} - \phi _{j-1,l + 1} )^{2n}  \right)
\right]\nonumber
\end{align}}\normalsize
where $N$ ($N\gg 1$) is the number of sites in our system, and
$m_0  = L(a_1  - 4a_2 )$. We introduce the following lattice
Fourier transform
\begin{equation}
\phi _{l,j}  = \frac{1}{{\sqrt A }}\sum\limits_{k_u,k_v } {\phi
(\vec k)e^{i(jk_u  + lk_v )r_0 } }
\end{equation}
where $A$ is total area of the lattice and $r_0$ is the lattice
spacing. We have chosen reciprocal lattice vectors corresponding
to the rhombic Bravais lattice defined by $\hat u$ and $\hat v$ .
Then $k_u$ and $k_v$ take on values which lie within the first
Brillouin zone for a rhombic lattice ($- \pi /r_0  < k_u  < \pi
/r_0$). This is not the only way we could choose our reciprocal
lattice; another possible choice corresponds to the first
Brillouin zone of the hexagonal lattice \cite{hexbrill}. However,
we have been able to show the equations we obtain do not depend on
this choice, and it is far easier to use the rhombic Brillouin
zone. Using Eq. (A4) we may rewrite our partition function as
\begin{equation}
Z = \int {D\phi (\vec k)} \exp \left( { - \frac{{E_0 [\phi (\vec
k)] + E_{anh} [\phi (\vec k)]}}{{k_B T}}} \right)
\end{equation}
where
\begin{align}
E_0  &= \frac{{Nm_0 }}{{2V}}\sum\limits_k {\phi (\vec k)} \left( 2
- 2\cos (k_a r_0 ))\right.\nonumber\\
&\left.+ (2 - 2\cos (k_b r_0 )) + (2 - 2\cos ((k_a - k_b )r_0 ))
\right)\phi ( - \vec k)
\end{align}
and{\scriptsize
\begin{align}
&E_{anh} [\phi (\vec k)] = L\sum\limits_{n = 2}^\infty
{\frac{{a_1 ( - 1)^{n - 1} + a_2 ( - 4)^n }}{{(2n)!}}}
\frac{N}{{V^n }}
\delta _{ - k_1 ,k_n  + k_{n - 1}  +  \ldots k_2 } \nonumber  \\
&\times \left. \left[ {\prod\limits_{m = 1}^{2n} {\sum\limits_{k_m
} {\phi (\vec k_m ) \left( {1 - e^{ - ik_{mu} r_0 } } \right)} } }
\right. + \prod\limits_{m = 1}^{2n} {\sum\limits_{k_m } {\phi
(\vec k_m )\left( {1 - e^{ - ik_{mv} r_0 } } \right)} }
\right.\nonumber\\
 &\left.+ \prod\limits_{m = 1}^{2n} {\sum\limits_{k_m }{\phi (\vec k_m
)\left({1 - e^{ - i(k_{mu}  - k_{mv} )r_0 } } \right)} }  \right]
\end{align}}\normalsize
where $\vec k_m  = \left( {k_{mu} ,k_{mv} } \right)$, thus
diagonalizing the 'free' part of the energy.

Let us consider the free energy for Gaussian fluctuations, where
we neglect $E_{anh}$. Here, we may integrate over $\phi(\vec k)$
and so obtain the free energy for $N \gg 1$:
\begin{align}
F_0  = 3NL(a_1  - a_2 )&\nonumber\\
 + \frac{k_B TNL}{2(2\pi )^2}\int_{ - \pi }^{\pi}  {dx} \int_{ - \pi }^{\pi}
 dy &\; \ln \left(
\frac{{m_0 }}{\pi }\left[ (2 - \cos x)\right. \right.\nonumber\\
+ &\left. \left.(2 - \cos y) + (2 - 2\cos (x - y)) \right]
\right).
\end{align}
The exact numerical calculation of the integral gives us
\begin{equation}
F_0  = 3NL(a_1  - a_2 ) + \frac{{k_B TNL}}{2}\ln m_0  + 0.235k_B
TNL.
\end{equation}
For the Gaussian correlation function, {\scriptsize
\begin{equation}
G_0 (\vec k) = Z^{ - 1} \int {D\phi (\vec k)} \phi (\vec k)\phi (
- \vec k)\exp \left( { - \frac{{E_0 [\phi (\vec k)] + E_{anh}
[\phi (\vec k)]}}{{k_B T}}} \right),
\end{equation}}\normalsize
it is easy to show that {\small
\begin{align}
&G_0 (\vec k) = {{k_B T}} \left[ m_0 \left( (2 - 2\cos (k_u r_0 ))\right. \right. \nonumber\\
&\left. \left. + (2 - 2\cos (k_v r_0 )) + (2 - 2\cos ((k_u  - k_v
)r_0 )) \right) \right]^{-1}.
\end{align}}\normalsize

To obtain the Hartree approximation we must develop perturbation
theory for the full correlation function. We will omit details how
one obtains the expansion \cite{diag}, but instead give the
Feynman rules that govern the perturbation expansions in our
theory. On expanding out $E[\phi (\vec k)]$we need three types of
diagrams (or vertices) to represent each $\varphi ^n$- term in
this expansion, for $n>2$ . For illustration some of the diagrams
are shown in Fig. 14.

\begin{figure}
\scalebox{1.1}{\includegraphics[3.5cm,24cm][5cm,25cm]{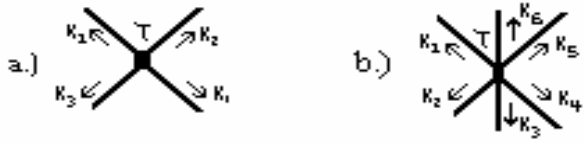}}
\vspace{1.5cm} \caption{a.) represents all three types of vertices
for the $\varphi^4$ term in the expansion for $E_{anh}$ . Shown,
here, are the four wave vectors associated with each $\varphi^4$
vertex. Also, we write an index $\tau = 1,2,3$ which distinguishes
between these three types. In the corresponding expression there
is a particular form factor associated with each of the three
types of vertex we may draw (see text). b.) represents all three
types of vertices for the $\varphi^6$  term. In general, a vertex
for the $\varphi^n$ term is obtained by drawing a point with lines
radiating from it. See text for a full discussion.} \end{figure}

Then, we may write down graphs to describe each term in our
expansion. Each graph will contain $N_V$ vertices, which are all
connected to each other by lines. We assign a label $i = 1 \ldots
N_V$ to each vertex.  For each vertex $i$, representing
$\varphi^n$, we must write down $\left( {a_1 ( - 1)^{n - 1}  + a_2
( - 4)^n } \right)$ $\delta _{ - k^i _1 ,k^i _n  + k^i _{n - 1} +
\ldots k^i _2 } /(2n)!$ and either one of the 'form' factors
{\small
\begin{align} \prod\limits_{m = 1}^{2n} {\left( {1 - e^{ -
ik_{mu}^i r_0 } } \right)} ,{\rm  }\prod\limits_{m = 1}^{2n}
{\left( {1 - e^{ - ik_{mv}^i r_0 } } \right)},\nonumber\\
\,\,\,\prod\limits_{m = 1}^{2n} {\left( {1 - e^{ - i(k_{mu}^i  -
k_{mv}^i )r_0} } \right)}
\end{align}}\normalsize
depending on whether the vertex is type 1, 2 or 3, respectively.
Each full Feynman graph will also consist of $N_E$  external lines
(connected to only one vertex) each associated with a wave vector
$\vec q_j$, $(j = 1, \ldots ,N_E )$. For each of these external
lines we write down $G(\vec q_j)$  and set  $\vec k_m^i  = \vec
q_j$ for one of the $\vec k_m^i$ in the vertex to which the line
is connected. There will also be $N_I$ internal lines, each
associated with a wave vector $\vec p_k$, $(k = 1, \ldots, N_I)$,
where each end is connected to two vertices, $i$ and $i'$. For
each of these internal lines we write down $G (\vec p_k)$ and set
$\vec k_m^i  = \vec k_m^{i'}  = \vec p_j$ for one of the $\vec
k_m^i$ in each of the two vertices. Then all the wave vectors for
the internal lines are summed over. Last of all, there is also a
symmetry factor that multiplies this, which accounts for how many
ways a term (graph) in the expansion may be generated.

To obtain the Hartree approximation we first consider only the
graphs for the full correlation function shown in Fig. 15. The sum
of these graphs we denote by $G_1$.
\begin{figure*}
\includegraphics[2.5cm,23cm][8cm,25.5cm]{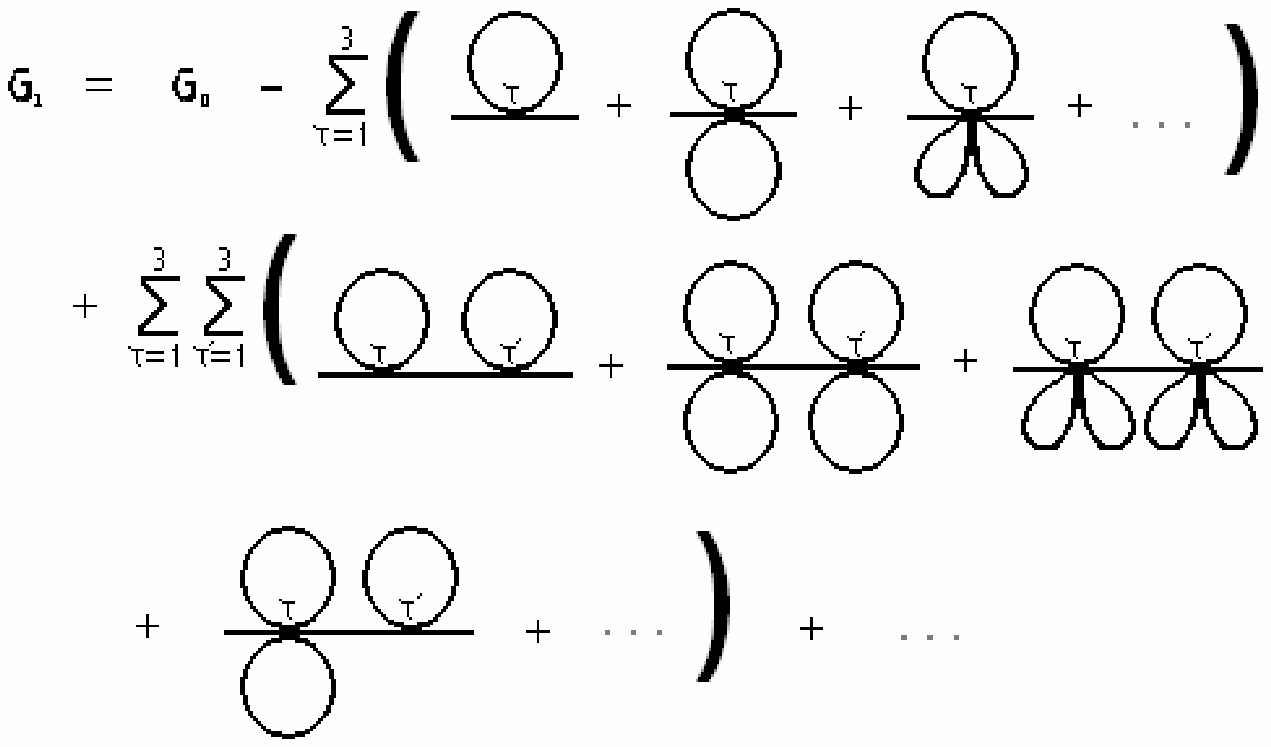}
\vspace{5cm} \caption{Graphs that are considered for the full
correlation function in the first stage of the Hartree
approximation.}
\end{figure*}
These form a series which we may easily sum
\begin{align}
&k_B TG_1 ^{ - 1} (\vec k) = M_1 (2 - \cos k_u r_0 ) \nonumber\\
&+ M_2 (2 - \cos k_v r_0 ) + M_3 (2 - \cos (k_u  - k_v )r_0 )
\end{align}
where
\begin{equation}
M_\tau   =  - a_1 L\sum\limits_{n = 0} {\frac{1}{{n!}}} \left(
{\frac{{k_B Tc_\tau  }}{{2m_0 }}} \right)^n  + 4a_2
L\sum\limits_{n = 0} {\frac{1}{{n!}}} \left( {\frac{{2k_B Tc_\tau
}}{{m_0 }}} \right).
\end{equation}
The zeroth order term in the series corresponds to $G_0$ . The
first order term corresponds to the $\varphi^4$  vertex, the
second order term to the $\varphi^6$ vertex and so on. For
$c_{\tau}$ we find the following expression:
\begin{align}
&c_\tau   = \frac{1}{{Nk_B T}}\sum\limits_{\vec p} {G(\vec p)}
\left[ (2 - 2\cos (k_u r_0 ))\delta _{\tau ,1} \right. \nonumber\\
&\left. + (2 - 2\cos (k_v r_0 ))\delta _{\tau ,2} + (2 - 2\cos
((k_u - k_v )r_0 ))\delta _{\tau ,3} \right].
\end{align}
From this we may show that $c_{\tau}=1/3$ and $M_1  = M_2  = M_3 =
m_1$. The next step is to go back to Fig. 15 and replace $G_0$ in
each loop with $G_1$, where $G_1$ will be replaced on the l.h.s.
of this expression with a new correlation function $G_2$. On
summing these graphs we find Eq. (A13) ($G_1$ replaced by $G_2$),
but with $M_1  = M_2  = M_3 = m_1$ where
\begin{equation}
m_2  =  - a_1 \exp \left( { - \frac{{k_BT}}{{6m_1 }}} \right) +
4a_2 \exp \left( { - \frac{{2k_BT}}{{3m_1 }}} \right).
\end{equation}
We then keep iterating this process until we have $J = m_\infty =
m_{\infty  - 1}$ and so obtain Eq. (6) of the text.

To calculate the free energy we must consider the sum of graphs
(Fig. 16).
\begin{figure*}
\includegraphics[2.5cm,24cm][5cm,25.5cm]{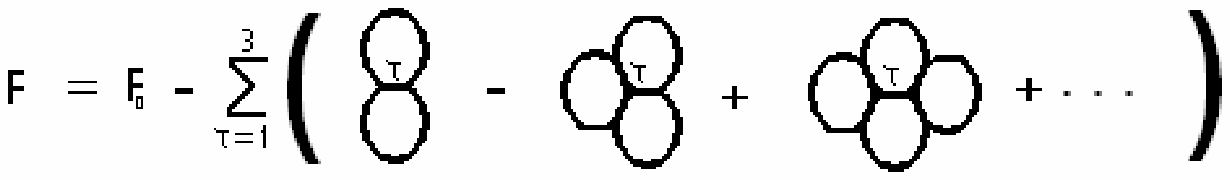}
\vspace{1.5cm} \caption{Diagrammatic expansion for the free
energy. Here, $F_0$ is the free energy calculated in the Gaussian
approximation.}
\end{figure*}
For the free energy in the Hartree approximation we take care in
replacing $m_0$ with $J$,  renormalizing our expansion, thus
obtaining Eq. (5) of the text.

\renewcommand{\theequation}{B\arabic{equation}}
\setcounter{equation}{0}
\section*{Appendix B: The Hartree approximation for the
quasi-antiferromagnetic state} \label{appb}

In the QAF state we now have spontaneous symmetry breaking where
\begin{align}
&\left\langle {\phi _{j,l}  - \phi _{j - 1,l} } \right\rangle  =
\left\langle {\phi _{j,l}  - \phi _{j - 1,l + 1} } \right\rangle =
\psi  \ne 0 \nonumber\\
&\mbox{and}\left\langle {\phi _{j,l}  - \phi _{j,l - 1} }
\right\rangle  = 0.
\end{align}
To take account of Eq. (B1) we rewrite $\phi _{j,l}  - \phi _{j -
1,l}  = \psi  + \phi '_{j,l}  - \phi '_{j - 1,l}$, $\phi _{j,l} -
\phi _{j - 1,l + 1}  = \psi  + \phi '_{j,l}  - \phi '_{j - 1,l +
1}$ and $\phi _{j,l}  - \phi _{j,l - 1}  = \phi '_{j,l}  - \phi
'_{j,l - 1}$. We then rewrite Eq. (A2) in the following form
{\scriptsize \begin{align} &E[\phi _{jl} ] = L\sum\limits_{p =
1}^2 {} \sum\limits_{jl} a_p (
- 1)^p \left[ [\cos (p(\phi '_{j,k}  - \phi '_{j,l - 1} )) \right. \nonumber\\
& - \cos(p\psi )(\cos (p(\phi '_{j,l}  - \phi '_{j - 1,l} )) +
\cos (p(\phi _{j,l}  - \phi _{j - 1,l + 1} )))] \nonumber\\
&\left. - [\sin (p\psi )(\sin (p(\phi '_{j,l} - \phi '_{j - 1,l}
)) + \sin (p(\phi '_{j,l}  - \phi '_{j - 1,l + 1} )))] \right]
\end{align}}\normalsize
Expanding in powers of  we may again divide the energy into
Gaussian and  anharmonic  terms:
\begin{equation*}
E[\phi _{j,l} ] = E_0 [\phi _{j,l} ] + E^{(1)} _{anh} [\phi _{j,l}
] + E^{(2)} _{anh} [\phi _{j,l} ];
\end{equation*}
\begin{align}
&E_0 [\phi _{j,l} ] = NL(a_2  - a_1 ) + 2NL(a_2 \cos (2\psi ) -
a_1 \cos (\psi ))\nonumber\\
&+ L(a_1 \sin (\psi ) - 2a_2 \sin (2\psi ))\Delta \phi '_{boundary} \nonumber\\
&+ \frac{1}{2}\sum\limits_{jl} \left[ {\tilde m_{0,1} (\phi
'_{j,l} - \phi '_{j - 1,l} )^2 }  + \tilde m_{0,2} (\phi '_{j,l} -
\phi '_{j,l - 1} )^2 \right. \nonumber\\
& \left. + \tilde m_{0,1} (\phi '_{j,l} - \phi '_{j - 1,l + 1} )^2
\right];
\end{align}
{\small
\begin{align}
 &E^{(1)} _{anh} [\phi _{j,l} ] = L\sum\limits_{jl} \sum\limits_{n = 2}^\infty  \frac{1}{{(2n)!}}\,  \left\{ \left( a_1 ( - 1)^{n - 1}  + a_2 ( - 4)^n  \right) \right. \nonumber\\ & \times \left( \phi '_{j,l}  - \phi '_{j,l - 1}  \right)^{2n}
 + \left( a_1 ( - 1)^{n - 1} \cos (\psi ) + \cos (2\psi )a_2 ( - 4)^n  \right) \nonumber\\
 &\left. \times \left( \left( (\phi '_{j,l}  - \phi '_{j - 1,l} )^{2n}  + (\phi '_{j,l}  - \phi '_{j - 1,l + 1} )^{2n}  \right) \right)
 \right\}\nonumber;
\end{align}
\begin{align}
  &E^{(2)} _{anh} [\phi _{j,l} ] = L\,\sum\limits_{jl} \sum\limits_{n = 2}^\infty
  \frac{1}{{(2n - 1)!}}  \left[ \left( a_1 \sin \psi ( - 1)^{n - 1}
  \right. \right.\nonumber\\
  &\left. + a_2 \sin (2\psi )( - 4)^n /2 \right)
   \left( (\phi '_{j,l}  - \phi '_{j - 1,l} )^{2n - 1} \right. \nonumber\\  &\left. \left. +
  (\phi '_{j,l}  - \phi '_{j - 1,l + 1} )^{2n - 1}  \right)
  \right]\nonumber
\end{align}}\normalsize
where
\begin{equation}
 \Delta \phi '_{boundary}  = \sum\limits_{jl} {((\phi '_{j,l} } -
\phi '_{j - 1,l} ) + (\phi '_{j,l}  - \phi '_{j - 1,l + 1} )),
\end{equation}
$\tilde m_{0,1}  = L(a_1 \cos \psi  - 4a_2 \cos 2\psi )$ and
$\tilde m_{0,2}  = L(a_1  - 4a_2 )$. Again, let us consider
Gaussian fluctuations and so discard all anharmonic terms. $\Delta
\phi '_{boundary}$ is the difference in $\phi '$  (fluctuations in
$\phi$) between one part of the boundary and another part of the
boundary of our lattice, which cannot be neglected. We can think
of the coefficient that multiplies $\Delta \phi '_{boundary}$ as a
net torque, which is non zero when the system is out of
equilibrium. At equilibrium, therefore, we require that, for the
$\Delta \phi '_{boundary}$ term to vanish,
\begin{equation}
\cos \psi  = \frac{{a_1 }}{{4a_2 }} \;\; \mbox{or} \; \sin \psi =
0.
\end{equation}
Indeed, through our definition of $\psi$; $\left\langle {\phi
'_{j,l}  - \phi '_{j - 1,l + 1} } \right\rangle$ ,  $\left\langle
{\phi '_{j,l}  - \phi '_{j,l - 1} } \right\rangle$ and
$\left\langle {\phi '_{j,l}  - \phi '_{j - 1,l} } \right\rangle$
must vanish, which occurs only if Eq. (B5) is satisfied. If only
the first condition of Eq. (B5) is satisfied, only fluctuations
around the antiferromagnetic state are under consideration. The
second condition corresponds to fluctuations in the ferromagnetic
state (c.f. Appendix A).

Now, we are able to perform the integrations over $\phi$ and so
arrive at the free energy
\begin{align}
  &F_0  = NL(a_1  - a_2 ) + 2NL(a_1 \cos (\psi ) - a_2 \cos (2\psi ))\nonumber \\
   &+ \frac{{k_B TNL}}
{{2(2\pi )^2 }}\int_{ - \pi }^\pi  {dx} \int_{ - \pi }^\pi  {dy}
\ln \left( {\frac{1}
{\pi }\left[ {\tilde m_{0,1} (2 - \cos x)} \right.} \right. \nonumber\\
 & \left. \left.
  + \tilde m_{0,2} (2 - \cos y) + \tilde m_{0,1} (2 - 2\cos (x - y)) \right]
  \right).
\end{align}
We shall leave the evaluation of this integral until later.

To obtain the Hartree approximation we proceed in the same manner
as in Appendix A and sum the same graphs for both the free energy
and the correlation function. Summing up all the graphs in Fig. 15
we find that Eqs. (A13) and (A14) still hold with $\tilde m_{0,1}$
replacing $m_0$, but now
 {\scriptsize
\begin{align}
c_\tau   &= \frac{1}{N}\sum\limits_{\vec p} {G(\vec p)} [(2 - 2\cos (k_u r_0 ))\delta _{\tau ,1}  \nonumber \\
&+ (2 - 2\cos (k_v r_0 ))\delta _{\tau ,2}  + (2 - 2\cos ((k_u  - k_v )r_0 ))\delta _{\tau ,3} ] \nonumber \\
&= \frac{1} {{(2\pi )^2 }}\int_{ - \pi }^\pi  dx\int_{ - \pi }^\pi
dy \times \nonumber\\ &\frac{{[(2 - 2\cos (x))\delta _{\tau ,1}  +
(2 - 2\cos (y))\delta _{\tau ,2}  + (2 - 2\cos (x - y))\delta
_{\tau ,3} ]}} {{(2 - 2\cos (x)) + \beta _0 (2 - 2\cos (y)) + (2 -
2\cos (x - y))}}
\end{align}}\normalsize
where $\beta _i  = \tilde m_{i,2} /\tilde m_{i,1}$ and $c_1  = c_3
\ne c_2$.

There is an important completeness relation $2c_1  + \beta c_2
\equiv 1$. We may evaluate one of the integrals and determine the
other through this relation. We find
\begin{align}
&c_1 (\beta ) = \frac{2} {\pi }\arcsin \left( {\frac{1} {{\sqrt
{2(\beta  + 1)} }}} \right)\nonumber \\
&c_2 (\beta ) = \frac{1} {\beta }\left( {1 - \frac{4} {\pi
}\arcsin \frac{1} {{\sqrt {2(\beta  + 1)} }}} \right).
\end{align}
So now we are able to set $M_1  = M_3  = \tilde m_{1,1}$ and $M_2
= \tilde m_{1,2}$ in Eq. (A13). Again, we go back to Fig. 15 and
replace $G_0$  in each loop with $G_1$,  and replace $G_1$ on the
l.h.s. with a new correlation function $G_2$.  So we have Eq.
(A13), but with $M_1  = M_3  = \tilde m_{2,1}$ and $M_2  = \tilde
m_{2,2}$ where
\begin{equation} m_{2,\tau }  =  - a_1 \exp \left( { - \frac{{kTc_\tau (\beta
_1 )}} {{2\tilde m_{1,0} }}} \right) + 4a_2 \exp \left( { -
\frac{{2kTc_\tau  (\beta _1 )}} {{\tilde m_{1,0} }}} \right).
\end{equation}

We again keep iterating this process until we have $J_1  =
m_{\infty ,1}  = m_{\infty  - 1,1}$ and $J_2  = m_{\infty ,2}  =
m_{\infty  - 1,2}$. Consequently, we obtain {\scriptsize
\begin{align}
J_1 &= L\left\{ {a_1 \cos \psi \exp \left( { - \frac{{k_B T}}
{{2J_1 }}c_1 (\alpha )} \right) - 4a_2 \cos 2\psi \exp \left( { -
\frac{{2k_B T}} {{J_1 }}c_1 (\alpha )} \right)} \right\}\nonumber\\
&J_2 = L\left\{ {a_1 \exp \left( { - \frac{{k_B T}} {{2J_1 }}c_2
(\alpha )} \right) - 4a_2 \exp \left( { - \frac{{2k_B T}} {{J_1
}}c_2 (\alpha )} \right)} \right\}
\end{align}}\normalsize
where $\alpha  = J_2 /J_1$. To obtain the free energy, again, we
must consider the sum of graphs shown in Fig. 16. To get the free
energy in the Hartree approximation we replace $\tilde m_{0,1}$
with $J_1$ and $\tilde m_{0,2}$ with $J_2$. Renormalizing the free
energy we find {\small
\begin{align}
F_{af}  &= \frac{{Nk_B T}} {2}\ln \left( {\frac{{J_1 }} {{k_B T}}}
\right) + \frac{{Nk_B T}} {6}\tilde \Omega _{af} (\alpha  +
\frac{1}{2}) + Nk_B T\hat C_{Hex}\nonumber \\
&+ 2NL\left[ a_0  - a_1 \cos \left( \psi \right)\exp \left(  -
\frac{{k_B T}} {{2J_1 }}c_1 (\alpha ) \right) \right. \nonumber\\
&\left. + a_2 \cos \left( {2\psi } \right)\exp \left(  -
\frac{{2k_B T}} {{J_1 }}c_1 (\alpha ) \right) \right] \nonumber\\
  &+NL\left[ a_0  - a_1 \exp \left(  - \frac{{k_B T}}
{{2J_2 }}c_2 (\alpha ) \right) \right.\nonumber \\
& \left.+a_2 \exp \left(  - \frac{{2k_B T}} {{J_2 }}c_2 (\alpha )
\right) \right]
\end{align}}\normalsize
where $\tilde \Omega _{af} (\alpha  + \frac{1} {2})$ is determined
through the following relationship
\begin{align}
  &\tilde \Omega _{af} \left( {\alpha  + \frac{1}
{2}} \right) = \frac{1} {{(2\pi )^2 }}\int_{ - \pi }^\pi  {dx}
\int_{ - \pi }^\pi  {dy} \times \nonumber\\& \ln \left(
{\frac{{\left[ {(2 - \cos x) + \alpha (2 - \cos y) + (2 - 2\cos (x
- y)} \right]}}
{{\left[ {(2 - \cos x) + (2 - \cos y) + (2 - 2\cos (x - y)} \right]}}} \right)\nonumber \\
   &= \int_1^\alpha  {d\alpha '} \frac{1}
{{\alpha '}}\left( {1 - \frac{4} {\pi }\arcsin \frac{1} {{\sqrt
{2(\alpha ' + 1)} }}} \right).
\end{align}
We could not find a closed form for this integral. Therefore, we
have approximated it with a logarithm and Pad\'{e} approximants
that interpolate between the Taylor expansion of $\tilde \Omega
_{af} (x)$ in powers of $x^{1/2}$ around $\tilde \Omega _{af}(0)$
and an asymptotic expansion.

Now, all the terms in $E^{(2)} _{anh} [\phi _{j,l} ]$ will
contribute with the requirement (from Eq. (B1)) that
{\scriptsize\begin{align}
\left\langle {\phi '_{j,l}  - \phi '_{j - 1,l} } \right\rangle  = \mathop {\lim }\limits_{p \to 0} (1 - \exp ( - iq_u r_0 )\Gamma (\vec q,\psi )) = 0 \nonumber\\
\left\langle {\phi '_{j,l}  - \phi '_{j - 1,l + 1} } \right\rangle
= \mathop {\lim }\limits_{p \to 0} (1 - \exp ( - i(q_u  - q_v )r_0
)\Gamma (\vec q,\psi )) = 0.
\end{align}}\normalsize
Eq. (B13) will constrain the value of $\psi$  as in the Gaussian
approximation. We find the following expression for $\Gamma (\vec
q,\psi )$ {\scriptsize\begin{align}
  &\Gamma (\vec q,\psi ) \equiv \nonumber\\ &\sum\limits_{n = 0}^\infty  {\frac{{a_1 \sin (\psi )\left( { - c_1 k_B T/(2\tilde m_{0,1} )} \right)^n  - 2a_2 \sin (2\psi )\left( { - 2c_1 k_B T/\tilde m_{0,1} } \right)^n }}
{{n!}}} \nonumber \\
   &= a_1 \sin (\psi )\exp \left( { - \frac{{k_B Tc_1 }}
{{2\tilde m_{0,1} }}} \right) - a_2 \sin (2\psi )\exp \left( { -
\frac{{2k_B Tc_1 }} {{\tilde m_{0,1} }}} \right).
\end{align}}\normalsize
Each term in this series can be represented diagrammatically as in
Fig. 17.
\begin{figure*}
\includegraphics[2.5cm,24cm][5cm,25.cm]{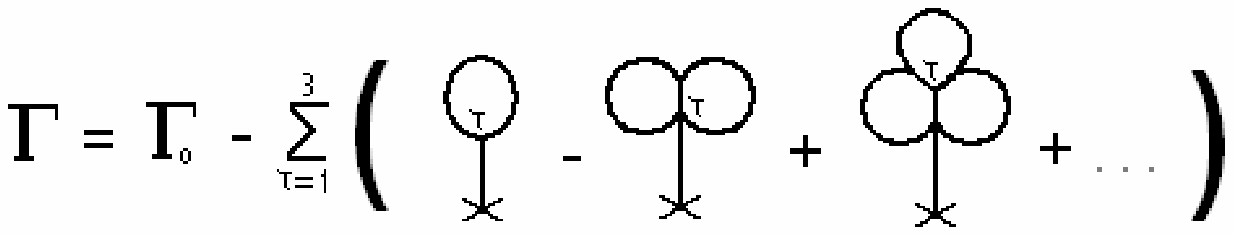}
\vspace{2cm} \caption{Diagrammatic expansion for $\Gamma (\vec
p,\psi )$. Here, $\Gamma _0  = a_1 \sin (\psi ) - 2a_2 \sin (2\psi
)$, and in the Gaussian approximation, $\Gamma (\vec p,\psi ) =
\Gamma _0$. Usual Feynman rules apply, except there is one
additional rule: the external line ending in a cross has no $G_0$
associated with it.}
\end{figure*}
To get the Hartree approximation, Eq. (B14) should be renormalized
by replacing $\tilde m_{0,1}$ with $J_1$ and $\beta$ with
$\alpha$, and so replacing $G_0$ by $G_{\infty}$ in each of the
loops. For Eq. (B13) to be satisfied we require that $\Gamma (\vec
p,\psi ) = 0$. This gives us the following equations for $\psi$:
\begin{equation}
\cos (\psi ) = \frac{{a_1 }} {{4a_2 }}\exp \left( {\frac{{3k_B
Tc_1 (\alpha )}} {{2J_1 }}} \right) \; \mbox{or} \; \sin(\psi)=0.
\end{equation}
The set of equations Eqs. (B15) , (B10) and (B11) form a complete
set for the QAF state.

For the QAF state on a hexagonal lattice close to $T=0$, from Eq.
(B5)  it follows that $\beta \le -1/2$. The upshot of this is that
$c_1(\beta)$ and $c_2(\beta)$ are complex unless $\beta = -1/2$
(at $a_1/4a_2=1$), and the QAF state at $T=0$ is unstable.
Therefore it cannot be a ground state except at the point of
frustration. At $T \ne 0$ we find that thermal fluctuations can
stabilize the QAF state. However, this state always has a higher
free energy than either the Potts or ferromagnetic states, so will
not be a phase of the system at thermal equilibrium.

If we allow the lattice to distort to the rhombic structure
described in the text, we have {\small\begin{align}
&E[\phi _{j,l} ] = L\sum\limits_{p = 1}^2 {} \sum\limits_{jl} {( - 1)^p [a_p (R_2 )\cos (p(\phi _{j,k}  - \phi _{j,l - 1} ))}\nonumber  \\
&+ a_p (R_1 )\cos (p(\phi _{j,l}  - \phi _{j - 1,l} )) + a_p (R_1
)\cos (p(\phi _{j,l}  - \phi _{j - 1,l + 1} ))].
\end{align}}\normalsize
On inspection of Eq. (B16), it is easy to modify Eqs. (B10), (B11)
and (B15) so as to arrive at Eqs. (11) and (12) of the text.

\renewcommand{\theequation}{C\arabic{equation}}
\setcounter{equation}{0}
\section*{Appendix C: The Hartree approximation for the Potts
state} \label{appc}

This is perhaps the most difficult of the states we must consider,
as so many of the lattice symmetries are broken. To perform
calculations we must make the construction shown in Fig. 18.
\begin{figure}
\scalebox{1.0}{\includegraphics[3.cm,24cm][6cm,25cm]{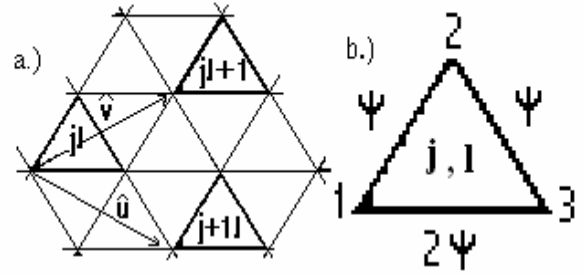}}
\vspace{3.5cm} \caption{a.) Lattice construction for the Potts
state where and now correspond to new lattice vectors for this new
unit cell. Here, we have constructed a new unit cell with the
basis shown in b.). Each of the three sites in the basis is given
a label ( 1, 2 or 3). This is to take into account the broken
lattice symmetries of the Potts state. Also shown is the magnitude
of the difference in $\left\langle \phi  \right\rangle$ between
each neighboring site, which may either be $\psi$ or $2\psi$.}
\end{figure}
Then, essentially, we must rewrite Eq. (A2)
\begin{align}
&E = E_1  + E_2  + E_3;\nonumber
\end{align}
\begin{align} &E_1  = \frac{1}
{2}\sum\limits_{p = 1}^2 {} \sum\limits_{j,l} {a_p ( - 1)^p } \times\nonumber\\
&\left[ {\cos (p(\phi _{1,j,l}  - \phi _{2,j,l} )) + \cos (p(\phi _{1,j,l}  - \phi _{2,j,l - 1} ))} \right. \nonumber\\
&+ \cos (p(\phi _{1,j,l}  - \phi _{2,j + 1,l - 1} )) + \cos (p(\phi _{1,j,l}  - \phi _{3,j,l} ))\nonumber \\
&\left. { + \cos (p(\phi _{1,j,l}  - \phi _{3,j,l - 1} )) + \cos
(p(\phi _{1,j,l}  - \phi _{3,j - 1,l} ))} \right];\nonumber
\end{align}
\begin{align}&E_2 = \frac{1}
{2}\sum\limits_{p = 1}^2 {} \sum\limits_{j,l} {a_p ( - 1)^p } \times\nonumber\\
&\left[ {\cos (p(\phi _{2,j,l}  - \phi _{3,j,l} )) + \cos (p(\phi _{2,j,l}  - \phi _{3,j-1,l + 1} ))} \right. \nonumber\\
&+ \cos (p(\phi _{2,j,l}  - \phi _{3,j - 1,l} )) + \cos (p(\phi _{2,j,l}  - \phi _{1,j,l} ))\nonumber \\
&\left. { + \cos (p(\phi _{2,j,l}  - \phi _{1,j-1,l + 1} )) + \cos
(p(\phi _{2,j,l}  - \phi _{1,j,l+1} ))} \right];\nonumber
\end{align} \begin{align} &E_3 = \frac{1}
{2}\sum\limits_{p = 1}^2 {} \sum\limits_{j,l} {a_p ( - 1)^p } \times\nonumber\\
&\left[ {\cos (p(\phi _{3,j,l}  - \phi _{1,j,l} )) + \cos (p(\phi _{3,j,l}  - \phi _{1,j,l + 1} ))} \right. \nonumber\\
&+ \cos (p(\phi _{3,j,l}  - \phi _{1,j + 1,l} )) + \cos (p(\phi _{3,j,l}  - \phi _{2,j,l} ))\nonumber \\
&\left. { + \cos (p(\phi _{3,j,l}  - \phi _{2,j+1,l} )) + \cos
(p(\phi _{3,j,l}  - \phi _{2,j+1,l-1} ))} \right];
 \end{align}
where we have introduced the angles $\phi_n$  where $n=1,2,3$,
according to their location on the lattice (the sites are defined
in Fig. 18). Then we must expand out this expression around the
Potts state, which we may do by writing $\phi _{1,j,l}  = \hat
\phi _{1,j,l}$, $\phi _{2,j,l}  = \psi  + \hat \phi _{2,j,l}$ and
$\phi _{3,j,l}  = 2\psi  + \hat \phi _{3,j,l}$. The expressions we
get are cumbersome and we shall not write them down. For brevity's
sake, in what follows we shall only give an outline of the
derivation and only state key results.

The next step is to separate $E$  into a Gaussian ($E_0$) and
anharmonic part ($E_{anh}$) in the same way as we did for the
other states. We may Fourier transform these to reciprocal space
using relations similar to Eq. (A4). Here we state a key result,
that
\begin{equation}
E_0  = \frac{1}{2}\sum\limits_{\vec k} \vec{\phi}  ^T ( - \vec
k)G_0^{ - 1} (\vec k)\vec{\phi} (\vec k)
\end{equation}
where $G_0^{-1}(\vec k)$ is given below (Eq. (C3)),
\begin{table*}[t]{\small\begin{align} G_0^{ - 1} (\vec k) =
\frac{{\tilde m_{0,1} }}{{k_B T}} \left( {\begin{array}{*{20}c}
   {3(1 + \beta _0 )} & { - 1 - e^{ - ik_v r_0 }  + e^{ - i(k_v  - k_u )r_0 } } & { - \beta _0 (1 + e^{ - ik_v r_0 }  + e^{ - ik_u r_0 } )}  \\
   { - 1 - e^{ik_v r_0 }  - e^{i(k_v  - k_u )r_0 } } & 6 & { - 1 - e^{ik_u r_0 }  - e^{i(k_u  - k_v )r_0 } }  \\
   { - \beta _0 (1 + e^{ik_v r_0 }  + e^{ik_u r_0 } )} & { - 1 - e^{ik_u r_0 }  - e^{i(k_u  - k_v )r_0 } } & {3(1 + \beta _0 )}  \\
\end{array}} \right)
\end{align}}\end{table*}
$\vec{\phi}  = \left( {\phi _1 ,\phi _2 ,\phi _3 } \right)$ and
$\beta _i = \tilde m_{i,2} /\tilde m_{i,1}$. We now define $\tilde
m_{0,1} = a_1 \cos (\psi ) - 4a_2 \cos (2\psi )$ and  $\tilde
m_{0,2}  = a_1 \cos (2\psi ) - 4a_2 \cos (4\psi )$. To obtain the
Hartree result for the correlation function we consider the same
graphs as in Fig. 15. Now, however, each $G_0$ is a $3 \times 3$
matrix, the inverse matrix of $G_0^{-1}$ given in Eq. (C3). On
summation of these graphs we find Eq. (C3), but with $G_0^{-1}$,
$\beta_0$ and $\tilde m_{0,1}$ replaced by  $G_1^{-1}$, $\beta_1$
and $\tilde m_{1,1}$, respectively, where $\tilde m_{1,1}$ and
$\tilde m_{1,2}$ are determined through the following relations
\begin{equation}
\tilde m_{1,s}  =  - a_1 \exp \left( { - \frac{{k_B T\eta _s
(\beta _0 )}}{{\tilde m_{0,1} }}} \right) + 4a_2 \exp \left( { -
\frac{{4k_B T\eta _s (\beta _0 )}}{{\tilde m_{0,1} }}} \right)
\end{equation}
where $s=$1 or 2, {\scriptsize\begin{align} &\eta _1 (\beta ) =
\frac{1}{{2(2\pi )^2 }}\int_{ - \pi }^\pi {dx} \int_{ - \pi }^\pi
{dy} \left[ (10 + 26\beta  + 6\beta ^2 ) \right. \nonumber\\ & -
(5 + 7\beta  + 2\beta ^2 )(\cos x + \cos(y))  - (8+ 10\beta +
2\beta ^2 )\cos (x - y) \nonumber\\ &\left. - \beta (\cos (x - 2y)
+ \cos (2x - y)) \right]G(x,y)^{ -
1}, \nonumber\\
&\eta _2 (\beta ) = \frac{1}{{(2\pi )^2 }}\int_{ - \pi }^\pi {dx}
\int_{ - \pi }^\pi {dy} \left[ 13 + 12\beta
- (4 + 6\beta )(\cos x + \cos y) \right. \nonumber\\
&  \left. { - 5\cos x\cos y - 3\sin y\sin x} \right]G(x,y)^{ -
  1}
\end{align}}
and
\begin{align}
 G(x,y;\beta )& = 6(6 + 13\beta  + 6\beta ^2 )\nonumber\\ &- 12(1 + \beta )^2 (\cos x + \cos y + \cos (x + y)) \nonumber\\
  &- 2\beta (\cos (x + y) + \cos (2x - y) + \cos (x - 2y)).
\end{align}

We may then calculate $G_2^{ - 1}$ as we did in the previous
expressions. We find again Eqs. (C3) and (C4), but with $G_1^{ -
1}$, $\tilde m_{0,1}$, $\tilde m_{0,2}$, $\tilde m_{1,1}$ and
$\tilde m_{1,2}$ all replaced by  $G_2^{ - 1}$, $\tilde m_{1,1}$,
$\tilde m_{1,2}$, $\tilde m_{2,1}$ and $\tilde m_{2,2}$,
respectively. On further iteration (as described in previous
appendices), we obtain Eqs. (8a) and (8b). Again we have used
Pad\'{e} approximants for $\eta_1 (\beta)$ and $\eta_2 (\beta)$.

In order to obtain the free energy again, we must consider the sum
of graphs shown in Fig. 16, but with each $G_0$ a matrix. Then we
replace $\tilde m_{0,1}$ with $J_1$ and $\tilde m_{0,2}$ with
$J_2$, obtaining Eq. 8 with
\begin{equation}
\tilde \Omega _{potts} (\alpha ) = \frac{1}{{(2\pi )^2 }}\int_{ -
\pi }^\pi  {dx\int_{ - \pi }^\pi  {dy} } \ln \left(
{\frac{{G(x,y;\alpha )}}{{G(x,y;1)}}} \right).
\end{equation}

$\Gamma (\vec k,\psi )$ is now a matrix, and is determined by
first considering the graphs shown in Fig. 17 (where $G_0$  is
also a matrix) and then replacing $\tilde m_{0,1}$ with $J_1$  and
$\beta$ with $\alpha$ . This enables us to derive Eq. (8c) of the
text.

\bibliographystyle{phaip}
\bibliography{mybib}

\begin{mcbibliography}{10}
\bibitem{top}
There are 6 distinct topological ground states corresponding to the broken
  symmetries of the Potts state. 3 states are got through translations of the
  lattice the other three are got by changing the staggered helicity (defined
  later in the text). However, what seems to matter in this transition is
  change in staggered helicity\relax
\relax
\end{mcbibliography}
%
%
%

\end{document}